\documentclass[12pt]{article}
\usepackage[ansinew]{inputenc} %damit Umlaute richtig übersetzt werden
\usepackage[T1]{fontenc} %für die DC-Fonts
\usepackage{amsmath}

\title{The Relativity of Discovery: \\
  Hilbert's First Note on the Foundations of
  Physics\thanks{Arch.~Hist.~Exact~Sci. 1998, in press.}}

\author{Tilman Sauer\\[0.5cm]
{\small Institut f\"ur Wissenschaftsgeschichte}\\ 
{\small Georg-August-Universit\"at G\"ottingen}\\
{\small Humboldtallee 11, 37073 G\"ottingen, Germany}\\
{\small e-mail: \texttt{tsauer@gwdg.de}}
}

\date{}

\begin{document}
\maketitle

Hilbert's paper on ``The Foundations of Physics (First
Communication),'' presented to the G\"ottingen Academy of Sciences on
November 20, 1915,%
\footnote{\cite{Hil1915Grundlagen1}.} is now primarily known for its
parallel publication of essentially the same gravitational field
equations of general relativity which Einstein published in a note on
``The Field Equations of Gravitation,'' presented to the Prussian
Academy of Sciences in Berlin five days later, on November 25, 1915.%
\footnote{\cite{EinsteinA1915Feldgleichungen}.}  An intense
correspondence between Hilbert and Einstein in the crucial month of
November 1915, furthermore, confronts the historian with a case of
parallel research and with the associated problem of reconstructing the
interaction between Hilbert and Einstein at that time.

Previous assessments of these issues have recently been challenged by
Leo Corry, J\"urgen Renn, and John Stachel who draw attention to a
hitherto unnoticed first set of proofs for Hilbert's note.%
\footnote{\cite{CorryLEtAl1997Decision}.} These proofs bear a
printer's stamp of December 6 and display substantial differences to
the published version, in particular as regards the covariance of the
theory and the discussion of the energy concept.  They also do not yet
contain the explicit form of Einstein's gravitational field equations
in terms of the Ricci tensor and its trace, the Riemann curvature
scalar.

By focussing on the consequences of these findings for the
reconstruction of Einstein's path towards general relativity --- any
possibility that Einstein took the clue for the final step towards his
field equations from Hilbert's note is now definitely precluded --- a
number of questions about Hilbert's role in the episode, however, are
left open. To what extent did Hilbert react to Einstein? What were
Hilbert's research concerns in his note, and how did they come to
overlap with Einstein's to some extent in the fall of 1915? How did
Hilbert and Einstein regard each other and their concurrent
activities at the time? What did Hilbert hope to achieve, and what,
after all, did he achieve?

With these questions in mind I shall discuss in this paper Hilbert's
first note on the ``Foundations of Physics,'' its prehistory and
characteristic features,%
\footnote{For the sake of completeness, it should be noted that, in
  1924, Hilbert published a revised version of his first note as the
  first part of a paper, titled also ``Die Grundlagen der Physik'' in
  the \emph{Mathematische Annalen} (\cite{Hil1924Grundlagen}). The
  second part of this 1924 note was his second note on the
  ``Foundations of Physics'' from December 1916
  (\cite{Hil1917Grundlagen2}), revised in 1924 as well. A discussion
  of this 1924 republication of his first note is beyond the scope of
  the present paper. It would involve a discussion of Hilbert's
  subsequent work leading to the publication of his second note, of
  later investigations of the energy problem by Felix Klein and
  others, as well as of developments in unified field theory up to
  1924.} and, for heuristic purposes, I shall do so largely from
Hilbert's perspective.

\section{Hilbert and his research program}

In contrast to Einstein's biography, Hilbert's life was curiously
unaffected by the drastic historical changes associated with Germany's
transition from the Kaiserreich to Weimar democracy to the Nazi
regime, or, for that matter, by the first World War. Hilbert came to
G\"ottingen in 1895, aged 33, as a young professor, and there he
stayed for nearly fifty years. In G\"ottingen he lived almost all
these years in the same house on Wilhelm-Weber-Stra{\ss}e 29, married
to the same woman, and leading a scholarly life devoted to mathematical
research and academic teaching. Hilbert never had a sabbatical, but he
gave lecture courses, term after term, on various fields of
mathematics, logic, and mathematical physics. These lectures are
documented by more than a hundred \emph{Vorlesungsausarbeitungen} kept
in the Mathematics Institute of G\"ottingen University. In spring and
fall, between terms, however, he used to leave G\"ottingen to
spend vacations either on the Baltic or in the Swiss mountains.%
\footnote{For biographical information on Hilbert, see
  \cite{BlumenthalO1935Lebensgeschichte} and \cite{ReidC1970Hilbert}.}

Intellectually, on the other hand, Hilbert's achievements have been
called revolutionary, in particular as regards his work on the
foundations of mathematics. And it would be wrong to assume that
Hilbert stewed in his own juice in G\"ottingen. He did participate
energetically in the scientific discussions of the day, and he also
took active part in matters of scientific interest to the small but
vital scientific community of G\"ottingen. It has often been said that
the people of G\"ottingen in the era of Klein and Hilbert saw the town
as one of the centers, if not as \emph{the} center, of the scientific
world.%
\footnote{For a discussion of the G\"ottingen mathematical community
  in the era of Klein and Hilbert, see \cite{RoweDE1989Klein}.}  Be
that as it may, Hilbert's active intervention in matters of science
policy aimed at having crucial people come to G\"ottingen.  Leading
scientists who worked creatively at the edge of current research in a
field that was of interest to G\"ottingen mathematicians were invited
to come there and inform Hilbert and his colleagues about
the latest developments in the field.%

A typical example is provided by Hilbert's first correspondence with
Einstein.%
\footnote{CPAE5, Docs. 378, 417.}  In 1912 Hilbert was working
on implications of his theory of linear integral equations for
physics, in particular the kinetic theory of gases and radiation
theory. This was a field where Einstein had published important
contributions. So Hilbert sent Einstein a postcard asking for
offprints of his papers.  Some time later he sent Einstein a copy of
his just published book on integral equations, and then invited him to
come to G\"ottingen in the spring of the following year to attend a
Wolfskehl symposium on the kinetic theory of matter.  Einstein
declined, saying he had nothing new to say and that he was also too
busy.

Intellectually, Hilbert's science policy is matched by a
characteristic feature of his perspective on the mathematical sciences
captured by the catchword ``the axiomatic method.'' The axiomatic
method is characteristically reflective; it takes up whatever insight
has been achieved into a field and tries to analyse, clarify and
reformulate it.  But it is not a vague, unspecific openness towards
anything new in the sciences. In physics, too, Hilbert had rather
firm judgments about important developments and the role he himself
and his group might play in them.

Hilbert's first note on the Foundations of Physics from 1915 indeed
represents the culmination of a lifelong interest and many years of
active work in physics.%
\footnote{For discussions of Hilbert's work in physics, see
  \cite{BornM1922Hilbert}, \cite{CorryL1996HilbertPhysics},
  \cite{CorryL1997Hilbert}.}  I would like to suggest that there are
some characteristics of his understanding of physics which date to his
early years and stay relatively invariant over the years.

\subsection{Physics as a mathematical discipline}

Hilbert's view of physics from a mathematician's perspective becomes
quite explicit in remarks he made regarding the relationship
between physics and geometry. Hilbert regarded geometry as a genuine
branch of mathematics. But, originally, geometry was a natural science.
Only it was no longer subject to experimental examination and had
become mathematized, arithmetized and eventually axiomatized. For
Hilbert, this development is not only an account of the factual
historical development but also of the proper advancement of science,
an advancement which should be furthered wherever possible.

Thus, as early as 1894, in a lecture on geometry which he gave while
still in K\"onigsberg, Hilbert wrote
\begin{quote}
  Geometry is a science which essentially has developed to such a state
  that all its facts may be derived by logical deduction from previous
  ones.%
  \footnote{ ``Die Geometrie ist eine Wissenschaft, welche im
    Wesentlichen so weit fortgeschritten ist, dass alle ihre
    Thatsachen bereits durch logische Schl\"usse aus fr\"uheren
    abgeleitet werden k\"onnen.'' SUB Cod.\ Ms.\ Hilbert 541, p.~7.}
\end{quote}
And he immediately adds
\begin{quote}
  Completely different from, e.g. electricity theory or optics where
  even today new facts are still being discovered.%
  \footnote{ ``Ganz anders wie z.B. die Electricit\"atstheorie oder
    Optik, in der noch heute immer neue Thatsachen entdeckt werden,''
    ibid.}
\end{quote}
Later in this lecture, in the course of discussing the axiomatic
foundations of geometry, he presents the axiom of parallels and
discusses the alternatives of Euclidean, hyperbolic and parabolic
geometries. In this context he remarks
\begin{quote}  
  Now also all other sciences are to be treated following the model of
  geometry, first of all mechanics, but then also optics and
  electricity theory.%
  \footnote{ ``Nach dem Muster der Geometrie sind nun auch alle
    anderen Wissenschaften in erster Linie Mechanik, hernach aber auch
    Optik, Elektrizit\"atstheorie etc.\ zu behandeln,'' ibid., p.~92.}
\end{quote}
This is 1894. Very similar remarks are made in a lecture course on
Euclidean geometry given in winter 1898/99. There Hilbert
characterizes geometry as
\begin{quote}
  a natural science but of such a kind that its theory may be called a
  perfected one which, as it were, provides a model for the
  theoretical treatment of other natural sciences.%
  \footnote{ `` [...] eine Naturwissenschaft [...], aber eine solche,
    deren Theorie eine vollkommene zu nennen ist, die gleichsam ein
    Muster bildet f\"ur die theoretische Behandlung anderer
    Naturwissenschaften.'' SUB Cod.\ Ms.\ Hilbert 551, p.~1.}
\end{quote}
In the same semester Hilbert also lectured on mechanics. This is in
fact Hilbert's first lecture course dealing with physics proper.%
\footnote{Except for a lecture course on hydrodynamics held in
 K\"onigsberg in summer 1887, cp.\ SUB Cod.\ Ms.\ Hilbert 522.}
In the introduction to this course, Hilbert again characterized geometry
as a mathematical science which used to be a natural science. Regarding
mechanics he then goes on:
\begin{quote}
  Also in mechanics the basic facts are accepted by all physicists.
  But the arrangement of the basic concepts nevertheless is subject to
  the changes in viewpoint. The structure is also far more complicated
  [than that of geometry]; even deciding what is simpler is something
  which depends on further discoveries. Hence, even today, mechanics
  cannot yet be called a purely mathematical discipline, at least not to
  the extent that geometry is.%
  \footnote{ ``Auch in der Mechanik werden die Grundthatsachen von
    allen Physikern zwar anerkannt. Aber die Anordnung der
    Grundbegriffe ist dennoch dem Wechsel der Auffassungen
    unterworfen. Auch der Aufbau ist ein viel complicirterer und was
    das einfachere ist, zu entscheiden ist von weiteren Entdeckungen
    abh\"angig, so dass die Mechanik auch heute noch nicht, jedenfalls
    nicht in dem Masse wie die Geometrie als eine rein mathematische
    Disziplin zu bezeichnen ist.'' SUB Cod.\ Ms.\ Hilbert 553, p.~2.}
\end{quote}
Given this state of affairs, Hilbert continues:
\begin{quote}    
    We must strive for it to become [a mathematical science]. We must
    extend the range of pure mathematics further and further, not only
    in our own mathematical interest but also for the sake of science
    as such.% 
\footnote{``Wir m\"ussen streben, dass sie es wird. Wir m\"ussen die Grenzen
    echter Mathematik immer weiter ziehen nicht nur in unserem
    mathematischen Interesse, sondern im Interesse der Wissenschaft
    \"uberhaupt.'' ibid.% 
}
\end{quote}

\subsection{Physics and the axiomatic method}

The lectures on Euclidean geometry and mechanics given in winter
1898/99 from which these quotes are taken immediately preceded the
writing of Hilbert's famous and classic ``Foundations of Geometry''
published in June 1899 as part of a \emph{Festschrift} on the occasion
of the unveiling of the Gauss-Weber-monument in G\"ottingen.%
\footnote{\cite{Hil1899Grundlagen}. The other part of this
  \emph{Festschrift} was a discussion by Emil Wiechert of the
  ``Foundations of Electrodynamics'' (\cite{WiechertE1899Grundlagen}).}
Hilbert's axiomatic treatment of geometry as laid out in the
\emph{Festschrift} was immensely influential in the emergence of what
is called the axiomatic method in mathematics, captured roughly by the
postulates of independence, consistency, and completeness for systems
of axioms, by the notion of implicit definition, and by the use of
models in the logical analysis of axiomatic systems.

Hilbert's work in geometry was also influential for his own
understanding of mathematics and physics. The quotes given above
illustrate that Hilbert saw his axiomatic reformulation of
Euclidean geometry as a model of the way in which physics was to be
treated as well. For a proper understanding of Hilbert's subsequent
work in physics some brief comments on the most explicit early
programmatic formulation for this work are in order: the sixth of the
23 problems for future mathematical research which Hilbert formulated
for the International Congress of Mathematicians in Paris in 1900,
i.e., roughly a year after the \emph{Festschrift}. This sixth problem
explicitly reads
\begin{quote}
  6. Mathematical Treatment of the Axioms of Physics.  The
  investigations on the foundations of geometry suggest the problem:
  \emph{To treat in the same manner, by means of axioms, those
    physical sciences in which already today mathematics plays an
    important part; in the first rank are the theory of probabilities
    and
  mechanics.}%
\footnote{``6. Mathematische Behandlung der Axiome der Physik.  Durch
  die Untersuchungen \"uber die Grundlagen der Geometrie wird uns die
  Aufgabe nahegelegt, \emph{nach die\-sem Vorbilde diejenigen
    physikalischen Disziplinen axiomatisch zu behandeln, in denen
    schon heute die Mathematik eine hervorragende Rolle spielt: dies
    sind in erster Linie die Wahrscheinlichkeitsrechnung und die
    Mechanik}.'' \cite[p.~272]{Hil1900Probleme}, translation, slightly
  adapted, from \cite[p.~454]{Hil1901/02Problems}.}
\end{quote}
The formulation of the sixth problem might suggest that Hilbert rather
specifically had in mind just the kind of logical analysis of basic
assumptions, derived theorems, and their mutual interdependence which
led to the axiomatic understanding of Euclidean geometry. His notes on
mechanics show, indeed, that Hilbert focussed on such logical elements,
for instance when he explicitly noted from the literature that not
only do Kepler's laws follow from Newton's law, but that some converse
assertion is also true. Namely, that if in the field of a central
force all motions are conic sections or, alternatively, all bounded
motions are closed, the force can only be the Newtonian $1/r^2$ or the
harmonic force $\propto r$. Hence, to some extent, Newton's law
conversely follows from Kepler's first law plus some additional assumptions.%
\footnote{SUB Cod.\ Ms.\ Hilbert 553, pp.~27f.}

Another such example is taken from Boltzmann's mechanics from 1897.%
\footnote{Ibid., p.~86. The example is taken from
  \cite[pp.~223--225]{BoltzmannL1897Vorlesungen}.} There Boltzmann
compares Gauss's dynamical principle of least action to the principle
of virtual displacements by looking at the motion of a material point
constrained to the convex side of a parabolic surface.  While Gauss's
principle allows the determination of the conditions when the point
leaves the surface, the principle of virtual displacements fails to do
so. In the latter case, additional assumptions have to be made.

These examples might suggest that Hilbert, in 1900, quite specifically
had in mind a reflection on the received knowledge of classical point
mechanics aiming at the identification of possible precise
formulations of its dynamical principles and their derivable
consequences. There is, however, another dimension to Hilbert's
program for the axiomatic understanding of physics. 
Hilbert, indeed, gives some further explanation of what he had in
mind. He mentions a number of ``important investigations by physicists
on the foundations of mechanics,'' namely the writings by Mach, Hertz,
Boltzmann, and Volkmann.%
\footnote{\cite[p.~272]{Hil1900Probleme},
  \cite[p.~454]{Hil1901/02Problems}. The references are to
  \cite{MachE1889Mechanik/2}, \cite{HertzH1894Prinzipien},
  \cite{BoltzmannL1897Vorlesungen}, and \cite{VolkmannP1900Einfuehrung}.}
Hilbert had studied Boltzmann's mechanics quite closely. He wrote:
\begin{quote}
  It is therefore very desirable that the discussion of the
  foundations of mechanics be taken up by mathematicians also. Thus
  Boltzmann's work on the principles of mechanics suggests the problem
  of developing mathematically the limiting processes, there merely
  indicated, which lead from the atomistic view to the laws of motion
  of continua. Conversely one might try to derive the laws of the
  motion of rigid bodies by a limiting process from a system of axioms
  depending upon the idea of continuously varying conditions of a
  material filling all space continuously, these conditions being
  defined by parameters. For the question as to the equivalence of
  different systems of axioms is always of great theoretical
  interest.%
  \footnote{``[...]; es ist daher sehr w\"unschenswert, wenn auch von
    den Mathematikern die Er\"orterung der Grundlagen der Mechanik
    aufgenommen w\"urde. So regt uns beispielsweise das Boltzmannsche
    Buch \"uber die Prinzipe der Mechanik an, die dort angedeuteten
    Grenzprocesse, die von der atomistischen Auffassung zu den
    Gesetzen \"uber die Bewegung der Continua f\"uhren, streng
    mathematisch zu begr\"unden und durchzuf\"uhren. Umgekehrt
    k\"onnte man die Gesetze \"uber die Bewegung starrer K\"orper
    durch Grenzprozesse aus einem System von Axiomen abzuleiten
    suchen, die auf der Vorstellung von stetig ver\"anderlichen, durch
    Parameter zu definierenden Zust\"anden eines den ganzen Raum
    stetig erf\"ullenden Stoffes beruhen --- ist doch die Frage nach
    der Gleichberechtigung verschiedener Axiomensysteme stets von
    hohem prinzipiellen Interesse.'' ibid.}
\end{quote}
The specification of the vague term ``mathematical treatment'' or
``axiomatic treatment'' is given here by referring to the problem of
a precise and rigorous mathematical formulation of the transition
between discrete, atomistic conceptions and continuum mechanics.  The
important point is Hilbert's concern with a proper mathematical
formulation of continuum mechanics.
 
In his early lectures, Hilbert always introduced a tripartite
division of mechanics: the mechanics of a single mass point, the
mechanics of systems of a finite number of mass points, and the
mechanics of infinitely many mass points.%
\footnote{See, e.g., SUB Cod.\ Ms.\ Hilbert 553, p.~7.}  In Hilbert's
understanding, the third part in itself was divided into continuum
mechanics which
did not only comprise elasticity theory and hydrodynamics but also the
investigation of the motion of those fluids which ``do not have all
the properties of matter,'' such as electric and magnetic fluids,
ether, and the motion of energy and entropy. The mechanics of infinitely
many mass points, on the other hand, also included molecular physics,
kinetic theory, the theory of ions and electrons as well as chemistry
in general.%
\footnote{See Hilbert's first lecture course on Continuum Mechanics
  from winter 1902/03 and summer 1903 as documented by an
  \emph{Ausarbeitung} by Berkowski kept at the Library of the
  Mathematics Institute of G\"ottingen University, p.~2; see also SUB
  Cod.\ Ms.\ Hilbert 553, pp.~145--147.}

A decision on the question whether matter ultimately consists of atoms
or whether it is some continuously extended substance is avoided.  One
of the major sources of Hilbert's knowledge of mechanics, however,
Kirchhoff's textbook from 1876, deals extensively with a continuum
mechanics under the explicit assumption that matter fills space
continuously, ``as it appears to do.''%
\footnote{``wie sie es zu thun scheint''
  \cite[p.~III]{KirchhoffG1877Vorlesungen1}. Hilbert's excerpts from
  Kirchhoff are kept in SUB Cod.\ Ms.\ Hilbert 516.}

The mechanics of infinitely many mass points (be it in its continuum
version or in its discrete, atomistic version), in any case, comprises
all of physics. The sixth problem, therefore, in some sense simply
formulates the task of mathematizing physics and of making it just one
more branch of mathematics.

Put this way, the Paris problem may seem preposterously ambitious. Not
necessarily so for Hilbert. He had solved some longstanding
mathematical problems by bringing together hitherto unrelated branches
of mathematics. His proof of the
finite basis theorem had put an abrupt end to decades of active
mathematical research in invariant theory by solving a central problem
on a higher level of abstraction, and a similar breakthrough was
achieved by his monumental \emph{Zahlbericht} which systematized and
summarized nineteenth-century knowledge in number theory.%
\footnote{See \cite{BlumenthalO1935Lebensgeschichte} and
  \cite{WeylH1944Hilbert} for a discussion of Hilbert's mathematical
  achievements.}  And, in the concluding passage of his Paris lecture,
Hilbert explicitly expressed a firm belief in the unity of the
mathematical sciences.
\begin{quote}
  Mathematical science is in my opinion an indivisible whole, an
  organism whose vitality is conditioned upon the connection of its
  parts. For with all the variety of mathematical knowledge, we are
  still clearly conscious of the similarity of the logical devices,
  the relationship of the ideas in mathematics as a whole and the
  numerous analogies in its different departments.%
  \footnote{``[...]; die mathematische Wissenschaft ist meiner Ansicht
    nach ein unteilbares Ganze, ein Organismus, dessen
    Lebensf\"ahigkeit durch den Zusammenhang seiner Teile bedingt
    wird. Denn bei aller Verschiedenheit des mathematischen
    Wissensstoffes im einzelnen gewahren wir doch sehr deutlich die
    Gleichheit der logischen Hilfsmittel, die Verwandtschaft der
    Ideenbildungen in der ganzen Mathematik und die zahlreichen
    Analogien in ihren verschiedenen Wissensgebieten.''
    \cite[p.~297]{Hil1900Probleme}, \cite[p.~478]{Hil1901/02Problems}.}
\end{quote}
Regarding the advancement of mathematical physics, Hilbert believed
that the mathematician's task was the deliberate application of
advanced and sophisticated mathematical methods and concepts. In his
Paris lecture, he mentioned Lie's theory of infinite transformation
groups as a possible means for systematically distinguishing systems
of axioms for physics.%
\footnote{\cite[p.~273]{Hil1900Probleme},
  \cite[p.~454]{Hil1901/02Problems}.} He also believed that the
calculus of variations was indeed just such a powerful, ``logical
device'' and that it could be utilized for a mathematically precise
formulation of
mechanics, in particular for continuous systems.%
\footnote{For a discussion of Hilbert's work in the calculus of
  variations, see \cite{ThieleR1997Variationsrechnung}; see also
  \cite{BlumP1994Bedeutung}.}  Another field of mathematical expertise
relevant for the advancement of mathematical physics was, of course,
the theory of both ordinary and partial differential equations.

So much for Hilbert's programmatic formulation of 20th century
mathematical physics. He was 38 and president of the \emph{Deutsche
  Mathematikervereinigung} when he delivered his address at the
International Congress of some two hundred mathematicians from all
over the world, an address which would establish his role as the
``Generaldirector'' of mathematical science, as Minkowski put it in an
enthusiastic letter to Hilbert.%
\footnote{\cite[p.~130]{MinkowskiH1973Briefe}.}  (Einstein, at the time
of Hilbert's Paris speech, was 21 years of age and was just spending a
few days of vacation with his mother and sister in the Swiss resort
village of Melchtal after having passed the oral examinations for his
ETH Diplom as ``Fachlehrer in mathematischer Richtung.''%
\footnote{CPAE1, Docs.~68--71.}
)

\section{Prehistory}

Hilbert, at that time, was a well-known if not yet world-famous mathematician.
In his five years as professor in G\"ottingen Hilbert had already had
a first occasion to decline a call to the university of Leipzig.
Another such occasion soon followed when Hilbert was offered the
prestigious chair of Lazarus Fuchs at Berlin University.%
\footnote{\cite[pp.~401, 406f]{BlumenthalO1935Lebensgeschichte}.}
Again he declined, but he succeeded in convincing the Prussian
ministry to create an extra professorship for mathematics which would
allow him to get Minkowski to come to G\"ottingen. His plan worked and
Minkowski came in 1902; and this is also the time, biographically,
when Hilbert again started to work on physics.  He begins with a
two-semester course on continuum mechanics. In the following semester
Hilbert and Minkowski also gave a joint seminar on the general problem
of the mechanical stability of continuous systems.  In this context,
Hilbert also started to do research, as witnessed by a talk to the
Naturforscherversammlung in Kassel in 1903. The talk was ``On Mechanics
of Continua'' and dealt with the problem of the mechanical stability
of a liquid in a vessel.%
\footnote{SUB Cod.\ Ms.\ Hilbert 593. The problem of the talk was how to
  control mathematically the infinitely many degrees of freedom if you
  investigate the stability of the liquid by looking at small
  deviations of its center of mass from its equilibrium position given
  that neither the shape of the surface nor its boundary conditions
  are fixed.}

\subsection{The period 1902--1915}

Notwithstanding Hilbert's early research work on continuum mechanics,
and notwithstanding a series of lecture courses and joint seminars on
the mathematical problems of mechanics, electromagnetism and other
fields of physics, G\"ottingen mathematical physics, as far as Hilbert
was concerned, was a task pursued predominantly by Minkowski.  Hilbert
in these years was working on his theory of integral equations. After
all, the axiomatization of physics was just one problem among 22
others on his list.

At that time, G\"ottingen was a stronghold of the so-called
electromagnetic world view.%
\footnote{See \cite{JungnickelCEtAl1986Mastery2}, pp.~227--245, for a
  discussion of the electromagnetic worldview in the setting of
  Germany's theoretical physics at the turn of the century.}
In 1902 and 1903, the G\"ottingen theoretician Max Abraham had
published programmatic papers on the dynamics of the electron.%
\footnote{\cite{AbrahamM1902Dynamik}, \cite{AbrahamM1902Prinzipien},
  \cite{AbrahamM1903Prinzipien}; for a historical discussion of
  Abraham's electron theory, see \cite{GoldbergS1977Abraham}.}  He had
shown that the inertial mass of the electron may completely be
accounted for on the basis of Maxwellian electromagnetism.

Assuming that the electron is a small rigid sphere with uniform charge
distribution, Abraham had computed the work to be done in changing the
energy and momentum of the electron's self-field when it is
accelerated by an external field.  Given that the electron was
regarded as the ultimate constituent of matter Abraham's result opened
up the perspective of a complete reinterpretation of the concepts of
Newtonian mechanics in terms of Maxwellian electrodynamics and the
foundation of physics on the basis of Maxwell's electromagnetic field
theory. Moreover, in doing the calculations, Abraham had obtained an
expression for the inertial mass of the electron which turned out to
depend on the electron's velocity.  And these results seemed to be in
good agreement with experimental data obtained by the G\"ottingen
experimental physicist Kaufmann on beta and cathode rays.%
\footnote{\cite{KaufmannW1902Masse}.} This confirmation of
Abraham's theory was ``certainly one of the most
important results of modern physics,'' as Lorentz observed in 1909.%
\footnote{\cite[p.~43]{LorentzH1909Theory}.}

Minkowski's subsequent role in the discussions of current theoretical
problems in electrodynamics and his work on special relativity and
four-dimensional electrodynamics%
\footnote{For recent historical discussions of Minkowski's work in
  relativity, see \cite{CorryL1997Minkowski} and
  \cite{WalterS1998Minkowski}.}  may have helped to effect a
reorientation of Hilbert's global perspective on physics. Hilbert,
too, gradually changed from his predominant concern with a
mathematically sound formulation of continuum mechanics to an
electromagnetic world view, which assigned some crucial importance to
the theoretical understanding of the electron.%
\footnote{Cp.\ the discussion in \cite{CorryL1998Hilbert}.}
 After Minkowski's death
and Born's work on the rigid body problem in special relativity,%
\footnote{\cite{BornM1909Dynamik}, \cite{BornM1909Masse}
  \cite{BornM1909Theorie}, \cite{BornM1910Kinematik}.}  it was two of Hilbert's students,
Erich Hecke and Wilhelm Behrens, who continued to work out details of
special-relativistic electron dynamics on the basis of the
Abraham-Born theory of the rigid spherical electron.%
\footnote{\cite{BehrensWEtAl1912Bewegung}.}

During this time, however, Hilbert himself worked predominantly on
mathematical problems, a major one being his theory of linear integral
equations. A decisive turn in Hilbert's research interests occurred in
1912, when he had just finished a major monograph on this subject,
summarizing the results and fruits of his research of the previous
eight years in a systematic exposition.%
\footnote{\cite{Hil1912Grundzuege}.} Working on the final chapters of
his book, he considered applications of his theory for physics and
discovered that it may provide a means of clarifying the foundations
of kinetic gas theory, and shortly later found that it also could be
applied to radiation theory. These insights called his old dreams of
mathematizing physics to life again. From now on, Hilbert devoted
his time during the next few years almost exclusively to mathematical
physics and after more than three years, in the fall of November 1915
his efforts promised to result in a major breakthrough, perhaps
comparable to his achievements in invariant theory or geometry: he
realized that a novel electromagnetic theory of the electron put
forward in 1912 by the Greifswald physicist Gustav Mie could be
incorporated into the same mathematical framework as another
completely novel theory of the gravitational field published just
recently by the young and rather creative theoretician Albert
Einstein. This would give Hilbert the chance of putting forward a
unified mathematical description of both the gravitational and
electromagnetic fields and, at the same time, a framework to compute
the dynamics of the electron and possibly to explain the quantum
features of the Bohr atom.

Hilbert's and Klein's science policy of systematically acquainting
themselves and their fellow G\"ottingen mathematicians with the most
recent research had paid off handsomely. Mie's theory was published in
three installments in 1912.%
\footnote{\cite{MieG1912Grundlagen1}, \cite{MieG1912Grundlagen2},
  \cite{MieG1913Grundlagen3}. For discussions of Mie's theory, see
  \cite[pp.~754--759]{PauliW1921Relativitaetstheorie},
  \cite[pp.~26--38]{VizginV1994Theories}.} But even before the last
communication was issued, Max Born gave a report on this theory to the
G\"ottingen mathematical society.%
\footnote{Born's report ``on the works by Mie on the theory of
  matter'' was given on December 17
  (JDM 22 (1913), p.~27), \cite{MieG1913Grundlagen3}
  was issued on December 31, 1912.} Born himself continued to work on
Mie's theory, reinterpreting it in terms of Minkowski's four-dimensional
formulation and working out formal relations to special-relativistic
elasticity theory. He gave another talk on Mie's theory and on some of
his own work a year later in late 1913.%
\footnote{Born gave a report on ``Mie's theory of matter'' to the
  G\"ottingen Mathematical Society on November 25, and lectured on
  some of his own work, ``carried out in connection with Mie's theory
  of matter,'' on December 16 (JDM 22 (1913), p.~207).
  \cite{BornM1914Impulsenergiesatz} was presented to the G\"ottingen
  Academy by Hilbert on December 20, 1913. Hilbert had also asked for
  offprints of Mie's papers in the fall of 1913, see Mie to Hilbert,
  22 October 1913, SUB Cod.\ Ms.\ Hilbert 254/1.} At that time, a
report on Einstein's and Grossmann's just recently published
\emph{Entwurf einer verallgemeinerten Relativit\"atstheorie und
  einer Theorie der Gravitation}
was given as well.%
\footnote{\cite{EinsteinAEtAl1913Entwurf} was published in June 1913
  (CPAE4, p.~340); a report on ``the work by Einstein and
  Grossmann on gravitation theory'' was given to the G\"ottingen
  Mathematical Society on December 9, 1913
  (JDM 22 (1913), p.~207).}  Einstein's theory with
its strange mathematical intricacies must have induced Hilbert to try
again to invite Einstein to G\"ottingen to learn about this theory at
first hand. In summer 1915 he succeeded,
and in the first week of July Einstein gave a course of six
Wolfskehl lectures on his general theory of relativity.%
\footnote{GM 1916, p.~13. Leo Corry first found notes on a part of
  Einstein's lectures taken by an unknown auditor and preserved in the
  Hilbert archives (SUB Cod.\ Ms.\ Hilbert 724).  These notes have
  meanwhile been published as Appendix B to
  CPAE6, pp.~586--590.  On June 29, 1915, Einstein lectured
  ``on gravitation'' at the G\"ottingen Mathematical Society
  (JDM 24 (1915), p.~68).}  Hilbert and Einstein were
enthusiastic about each other on this their first encounter.%
\footnote{\label{note:EHcorrespondence}For Einstein's reaction to
  Hilbert, see Einstein to Zangger, 7 July 1915, CPAE8, Doc.~94;
  Einstein to Sommerfeld, July 15, 1915
  (\cite[p.~30]{HermannA1968EinsteinSommerfeld}) CPAE8, Doc.~96;
  Einstein to W.\ and G.\ de~Haas, August 1915
  (\cite[p.~259]{PaisA1982Subtle}) CPAE8, Doc.~110. For Hilbert's
  reaction, see the quotes given below.}  At the end of the summer
term, Hilbert was left with a number of rather interesting but
intricate papers on mathematical physics which called for competent
mathematical analysis. He would soon find not only exciting analogies
and connections between those two theories but also pinpoint flaws in
Einstein's rather pedestrian way of dealing with the mathematics of
his gravitation theory.

\subsection{Mid-July to Mid-November 1915} 

Unfortunately, very little is known about Hilbert's whereabouts and
intellectual preoccupations in the late summer and fall of 1915 after
Einstein's visit. One of the last documents from this summer is a
letter to Schwarzschild, dated 17 July.  In this letter Hilbert wrote
\begin{quote}
  We had a lot of scientific business here during the two war
  semesters. Almost all mathematics and physics lecturers incl.\
  Voigt and Tammann participate in my seminar; the main instigator
  is, of course, Debye.  During the summer we had as guests one after
  the other: Sommerfeld, Born, Einstein. In particular, the lectures
  of the latter on gravitation theory were memorable.%
\footnote{``Wir haben hier w\"ahrend der beiden Kriegssemester vielen
  wissenschaftlichen Betrieb gehabt. An meinem Seminar nehmen fast
  alle math.\ u.\ phys.\ Dozenten incl.\ Voigt u. Tammann teil; der
  Hauptmacher ist nat\"urlich Debye. W\"ahrend des Sommers hatten wir
  hier zu Gast der Reihe nach: Sommerfeld, Born, Einstein. Besonders
  die Vortr\"age des letzteren \"uber Gravitationsth.\ waren ein
  Ereigniss.'' 
  Hilbert to Schwarzschild, July
  17, 1915, SUB Cod.\ Ms.\ Schwarzschild 331/7.}
\end{quote}
Four days later, a meeting in preparation for Emmy Noether's
habilitation was held in G\"ottingen, and it is known that Hilbert,
the most fervent proponent and advocate of Noether's habilitation,
attended.%
\footnote{\label{note:Tollmien}\cite[p.~15]{TollmienC1991Habilitation}.
  Tollmien gives a detailed historical account of Emmy Noether's
  habilitation with extensive quotes of the extant archival
  documents.}  Hilbert's presence was also needed since the affair was
more than routine academic business.  It would set a precedent and
implied breaking a decree issued in 1908 by the Prussian ministry
against the habilitation of women at Prussian universities.%
\footnote{As it turned out this first attempt eventually failed.  Emmy
  Noether's habilitation was only achieved in G\"ottingen in 1919
  after the end of the war and after the German November revolution,
  see \cite{TollmienC1991Habilitation}.}  After this July meeting
it is not known what Hilbert did or where he was.

Then there is a postcard from Hilbert's student Richard B\"ar who later
worked out Hilbert's relativity lectures, dated August 17, 1915.%
\footnote{SUB Cod. Ms. Hilbert 772/1.} B\"ar had visited Hilbert in
Switzerland and the two sent greetings to Hilbert's wife who had
remained in G\"ottingen.

Six years later, Felix Klein reports in a letter to Pauli that Hilbert
had the decisive insight (``die entscheidende Gedankenwendung'') in %
the fall of 1915 in R\"ugen, and adds that Sommerfeld should know more
about this.%
\footnote{Klein to Pauli, 8 Mai 1921,
  \cite[p.~31]{PauliW1979Briefwechsel}.}  Klein's recollection is
corroborated by a letter from Erich Hecke to Hilbert, dated 16 October,
in which he thanks him for a postcard ``von R\"ugen aus.''%
\footnote{SUB Cod. Ms. Hilbert 141/5.} While it is thus probable that
Hilbert did spend some time during the fall of 1915 on R\"ugen, we 
don't know exactly when.%
\footnote{The question may be relevant because Einstein also spent a
  couple of weeks in R\"ugen in the second half of July, i.e., after he
  gave his talks in G\"ottingen. There is hence a slight chance that
  Hilbert was in R\"ugen at the same time as Einstein and that the two
  continued their discussion there.  However, as far as I know there
  is no further indication in the correspondence that would
  corroborate this surmise.}

The next thing we know about Hilbert is that he spent a week in Munich
in mid-October. On his way back, he met Schwarzschild's mother in the
train, and took the occasion to write a postcard to Schwarzschild, dated
Oct. 23:
\begin{quote}
  Dear S[chwarzschild].\\
  Just as we were coming back from an eight-day visit to Munich we met
  your mother in the train.---During the summer semester we had many
  scientific guests in G\"ottingen, among them also Einstein, [which]
  was highly interesting. The ``astronomical'' Freundlich was with us as
  well.  The astronomers, I think, must now leave everything else
  aside and
  only try to verify or refute Einstein's gravitation law!\\
  Yours sincerely.%
  \footnote{ ``Lieber S[chwarzschild].  Eben von einem 8t\"agigen
    Ausflug von M\"unchen kommend, treffen wir Ihre Frau Mutter im
    Zuge.---Im Sommersemester haben wir in G\"ottingen viele
    wissenschaftliche G\"aste gehabt, darunter auch Einstein, war
    hochinteressant. Auch der astronomische Freundlich war bei uns.
    Die Astronomen, meine ich, m\"ussen nun Alles liegen lassen u.\ nur
    darnach trachten das Einsteinsche Gravitationsgesetz zu
    best\"atigen oder widerlegen!  Haben Sie herzlichste
    Gr\"u{\ss}e.'' Hilbert to Schwarzschild, October 23, 1915, SUB
    Cod.\ Ms.\ Schwarzschild 331/6; see also Pringsheim to Hilbert, 25
    October 1915, SUB Cod.\ Ms.\ Hilbert 318/6.}
\end{quote}
With his return to G\"ottingen, a rather busy time began for Hilbert
with the daily routine of the winter term. A first Academy meeting
which always took place on Saturdays was held on October 23.%
\footnote{GAA Chron 2.1, Vol.~6, 410.} Then he had to give his
lectures: Mondays he lectured for two hours on differential equations
and gave a two-hour seminar on the structure of matter in the
afternoon together with Debye.%
\footnote{\cite[pp.~14, 17]{Verzeichnis1915WSVorlesungen}. An
  \emph{Ausarbeitung} of his lecture course on differential equations
  is kept at the Mathematics Institute of G\"ottingen University.} On
Monday
evenings he attended the physics colloquium.%
\footnote{See Hilbert to Einstein, 13 November 1915, CPAE8, Doc.~140.}
Tuesdays meetings of the G\"ottingen Mathematical Society were held.%
\footnote{The meetings started on November 2 with a session on
  ``Ferienberichte,'' JDM 24 (1915), p.~111.}  On
Thursday mornings another two hours of lectures on differential
equations were scheduled, and in the afternoon there were the weekly
hikes of Hilbert,
Klein and Runge where science policy matters had to be discussed.%
\footnote{See, e.g., \cite[p.~101]{Hil1909Minkowski},
  \cite[p.~407]{BlumenthalO1935Lebensgeschichte},
  \cite[p.~123]{RungeI1949Runge}.}

On the following Friday, October 29, a commission meeting concerning the
habilitation of Emmy Noether took place. Another meeting of the
Mathematics-Physics Department was scheduled for the following week, on
Saturday, November 6. On Tuesday, November 9, Emmy Noether gave a talk
on transcendental numbers to the G\"ottingen mathematical society.%
\footnote{JDM 24 (1915), p.~111,
  \cite[p.~14]{DickA1970Noether}. Dick quotes a letter by Emmy Noether
  to Fischer: ``Das hat sich sogar der hiesige Geograph angeh\"ort,
  f\"ur den es ein bi{\ss}chen sehr abstrakt war; die Fakult\"at will
  sich in ihrer Sitzung von den Mathematikern keine Katze im Sack
  verkaufen lassen.'' ibid.} On the day after that, the
Historical-Philosophical Department discussed Noether's petition for
habilitation; these colleagues were particularly conservative.%
\footnote{\cite{TollmienC1991Habilitation}, cp.\ note
  \ref{note:Tollmien}.}

Despite all these matters to be seen of, Hilbert still tried to
work out the implications of his ``entscheidende Gedankenwendung,'' as
Klein had called it, implying generally covariant field equations for
the gravitational and for the electromagnetic fields, some
mathematical relations between these field equations, a discussion of
the energy theorem in this framework, and the prospect of finding a
solution to the non-linear generalized Maxwell equations that would
correctly describe the electron. However, on the previous weekend
Hilbert had received an alarming letter from Einstein, who had sent him
the proofs of his first November communication,%
\footnote{\cite{EinsteinA1915Relativitaetstheorie}.}  which introduced
Einstein's return to general covariance%
\footnote{There is an extensive literature on Einstein's path towards
  General Relativity, see, in particular the classical accounts of
  \cite{StachelJ1980Search} and \cite{NortonJ1984Einstein}. See also
  \cite{RennJEtAl1998Heuristics} and the references cited therein.}
and also presented a modified set of gravitational field equations
that, however, were not yet generally covariant. Einstein wrote:
\begin{quote}
  With the same mail I am sending you the proofs of a paper in which I
  modified the gravitation equations after I myself had realized some
  four weeks ago that my former proof was flawed. Colleague Sommerfeld
  wrote to me that you also found a hair in my soup which made it
  unpalatable to you. I am curious whether you will come to like this
  new solution.%
  \footnote{``Mit gleicher Post sende ich Ihnen die Korrektur einer
    Arbeit, in der ich die Gravitations\-gleichungen abge\"andert habe,
    nachdem ich selbst vor etwa 4 Wochen erkannt hatte, dass mein
    bisheriges Beweisverfahren ein tr\"ugerisches war. Kollege
    Sommerfeld schrieb mir, dass auch Sie in meiner Suppe ein Haar
    gefunden haben, das sie Ihnen vollkommen verleidete. Ich bin
    neugierig, ob Sie sich mit dieser neuen L\"osung befreunden
    werden.'' Einstein to Hilbert, 7 November 1915, CPAE8, Doc.~136.}
\end{quote}
Hilbert responded with a friendly letter, now lost. But matters got
worse on the following weekend when he received a second letter from
Einstein%
\footnote{Einstein to Hilbert, 12 November 1915, CPAE8, Doc.~139.}  in which
Einstein reported further progress on achieving general covariance in
his theory of gravitation. In his second
communication to the Berlin Academy,%
\footnote{\cite{EinsteinA1915Nachtrag}.} Einstein now had published
generally covariant field equations using the Ricci tensor but at the
price of assuming that the trace of the energy-momentum tensor vanish
as it does for the electromagnetic energy-momentum tensor. In his
letter Einstein alluded to this consequence by pointing out that, by
this hypothesis, gravitation must play a fundamental role in the
constitution of matter.

But this touched on what Hilbert saw as his own original insight.
Alarmed, Hilbert decided to take action. He announced a lecture on the
``fundamental equations of physics'' in the G\"ottingen Mathematical
Society where he wanted to present his own recent investigations.%
\footnote{JDM 24 (1915), p.~111: ``Dritte Sitzung am
  16.~November. Hilbert, Grundgleichungen der Physik.''} 
He called Emmy Noether for assistance. She wrote to Fischer in Erlangen
\begin{quote}
  Invariant theory here is now trump; even [Paul] Hertz, the physicist,
  studies Gordan-Kerschensteiner; Hilbert wants to talk next week
  about his Einsteinian differential invariants, and so the
  G\"ottingers must get up to speed.%
  \footnote{\label{note:Kerschensteiner}``Invariantentheorie ist jetzt
    hier Trumpf; sogar der Physiker [Paul] Hertz studiert
    Gordan-Kerschensteiner; Hilbert will nächste Woche \"uber seine
    Einsteinschen Differentialinvarianten vortragen, und da m\"ussen
    die G\"ottinger doch etwas k\"onnen.''
    \cite[p.~14]{DickA1970Noether}.  Unfortunately, Dick does not give
    a source for this letter and dates it only unspecifically to
    November 1915. The textbook on invariant theory referred to is
    \cite{KerschensteinerG1885Vorlesungen},
    \cite{KerschensteinerG1887Vorlesungen}. In summer 1915 Hertz
    corresponded with Einstein about the latter's hole argument; see
    \cite{HowardDEtAl1993Labyrinth}, for a discussion of this
    correspondence.}
\end{quote}
And on Saturday, November 13, he sent two postcards to Einstein asking
him to come to his lecture on the following Tuesday, November 16, or
better still to the physics
colloquium on Monday evening, inviting him to stay at Hilbert's house.%
\footnote{Hilbert to Einstein, 13 November 1915, CPAE8, Doc.~140.}
Einstein declined.%
\footnote{Einstein to Hilbert, 15 November 1915, CPAE8, Doc.~144.}  He was, in
fact, just then working out what proved to be perhaps his greatest
breakthrough, the correct computation of the perihelion advance of
Mercury on the basis of his new field equations.

So Hilbert had to report his findings in correspondence to Einstein,
unfortunately lost. He probably sent Einstein the manuscript of
his lecture to the G\"ottingen Mathematical Society, or a summary of
its main points.%
\footnote{In his letter of November 18, Einstein responded to this
  lost piece of correspondence, that, as far as Einstein could tell,
  Hilbert's system was equivalent to the one he, Einstein, had
  found in the preceding weeks and presented to the Berlin Academy
  (Einstein to Hilbert, 18 November 1915, CPAE8, Doc.~148).}  But he realized
that he did not have any more time to carefully work out the
consequences of his own theory as he had hoped to do, in particular
that he would not succeed in obtaining a solution to the electron
problem in time.  The next occasion for presenting a communication to
the G\"ottingen Academy was Saturday, November 20, but Hilbert had
already sent his manuscript on the ``fundamental equations of
physics'' to the printer a day ahead, on November 19, a rather
unusual irregularity in the Academy's
proceedings.%
\footnote{\label{note:Journal} The invitation for the meeting of 20
  November was issued on November 15 and was, as always, circulated
  among the members to confirm their participation and announce any
  communications they intended to present at the meeting. Into this
  invitation, Hilbert wrote: ``Hilbert legt vor in die Nachrichten:
  Grundgleichungen der Physik.'' (GAA Chron 2.1, Vol.~6, 411). For
  each communication to the ``Nachrichten'', the ``Journal f\"ur die
  `Nachrichten' der Gesellschaft der Wissenschaften in G\"ottingen,''
  (GAA Scient~66, Nr.~2), lists author and title of the paper, the
  member of the Academy who presented it for publication, and the date
  of presentation.  In addition, there are columns ``To the printer''
  (``Zum Druck''), ``correction'' (``Korrektur''), and an untitled
  column which has the information about the issue in which the
  article was finally published.  For Hilbert's communication, entry
  Nr.~730 of the ``Journal,'' the title ``Die Grundgleichungen der
  Physik (Erste Mitteilung)'' was later corrected to ``Die Grundlagen
  der Physik (Erste Mitteilung).''  In the period covered by the
  ``Journal'', 1912--1935, there were eight items out of several
  hundred which were given to the printer \emph{before} being
  presented to the Academy.} And he must have been glad he did, since
on that day he received notice from Einstein that he had succeeded in
calculating the correct perihelion advance.%
\footnote{Einstein to Hilbert, 18 November 1915, CPAE8, Doc.~148. On that day
  Einstein had presented his third communication
  (\cite{EinsteinA1915Erklaerung}) to the Berlin Academy. See
  \cite{EarmanJEtAl1993Explanation} and CPAE4, pp.~344--359, for a
  discussion of Einstein's paper on the perihelion advance of Mercury
  and for a reconstruction of the rapidity of his calculations.}  To
that Hilbert could only respond with a ``herzliche Gratulation,'' and
the remark
\begin{quote}
  If I could do the calculations as rapidly as you, the electron would
  have to surrender and the hydrogen atom would have to produce a
  letter from home excusing it from not radiating.%
  \footnote{``Wenn ich so rasch rechnen k\"onnte, wie Sie, m\"usste
    bei meinen Gleichg das Elektron kapituliren und zugleich das
    Wasserstoffatom sein Entschuldigungszettel aufzeigen, warum es
    nicht strahlt.'' Hilbert to Einstein, 19 November 1915, CPAE8,
    Doc.~149.}
\end{quote}

\section{The proofs and the published paper}

Hilbert's note was published under the title ``The Foundations of
Physics (First Communication)'' in the last issue of the 1915 volume
of the \emph{Nachrichten} of the G\"ottingen Academy of Sciences.%
\footnote{``Die Grundlagen der Physik (Erste Mitteilung)''
  \cite{Hil1915Grundlagen1}. The published version was issued on 31
  March 1916 (\cite[p.~1271]{CorryLEtAl1997Decision}) but apparently
  Hilbert had received offprints or proofs of the final version by
  mid-February 1916 which he sent to various colleagues, see, e.g.,
  P.~Hertz to Hilbert, 17 February 1916, SUB Cod.\ Ms.\ Hilbert 150/2,
  L.~K\"onigsberger to Hilbert, 20 February 1916, SUB Cod.\ Ms.\ 
  Hilbert 187/13, G.~Mie to Hilbert, 13 February 1916, SUB Cod.\ Ms.\ 
  Hilbert 254/2. These offprints, however, apparently did not yet bear
  the pagination of the published version but had page numbers 1ff.,
  see Hilbert to Schwarzschild, 26 February 1916, SUB Cod.\ Ms.\ 
  Schwarzschild 331/8. \cite{KleinF1917Note} als refers to page
  numbers 1ff.\ The page references in Klein's note were updated to
  the page numbers 395ff.\ in the reprint of his Collected Papers.
  \cite[p.~810]{EinsteinA1916Grundlage} referred to a p.~3 of Hilbert's
  note.} As mentioned above, attention has recently been drawn
to a first set of proofs for this note.% 
\footnote{\cite{CorryLEtAl1997Decision}. The proofs are preserved in
  SUB Cod.~Ms. Hilbert 634 (in the following referred to as
  ``proofs''). They consist of 13 pages of consecutive pagination.
  Pages are printed on both sides.  Pp.~1/2, [7/8], and 13 are on a
  single sheets, pp.~3/4 and 5/6 as well as pp.~9/10 and 11/12 are on
  folded signature sheets. From the single sheet which contains
  pp.~[7/8] a piece was cut off from the top such that approximately
  10 lines are missing on the top of these pages including the page
  numbers. Pp.~1, 9, and 13 display a printer's stamp by the
  ``Dieterichsche Univ.-Buchdruckerei W.~Fr.~Kaestner G\"ottingen''
  with the date of December 6, 1915. On the top of p.~1 Hilbert added
  in dark ink ``Erste Korrektur meiner ersten Note.''  Underneath
  these words, some words had been written which were later erased.
  The pages contain some correction marks but also some marks in
  pencil which were later erased as well. On the top right corners of
  pp.~1, 3, and 7, Roman numbers I, II, III were added in ink. It is
  apparently these three sheets which Hilbert sent to Klein in March
  1918, see Hilbert to Klein, March 7, 1918
  (\cite[p.~144]{Hil1985EtAlBriefwechsel}).}  This first set bears a
printer's stamp of December 6 and displays substantial differences
from the published version, in particular with regard to the
discussion of the concept of energy and the covariance of the theory.
In contrast to standard accounts of the history of General Relativity
it was pointed out that, in this first version, Hilbert still believed
in Einstein's argument for the necessity of distinguishing between
what Hilbert called world parameters and space-time coordinates, i.e.,
the necessity of introducing the so-called adapted coordinate systems
of Einstein's \emph{Entwurf} theory, which restrict the general
covariance of the theory. Corry, Renn, and Stachel, in particular,
emphasize the fact that the proofs do not yet contain the explicit
form of the gravitational field equations in terms of the Ricci tensor
and the Riemann curvature scalar.

Despite these important differences, however, there are also a number
of characteristic features of Hilbert's theory as presented in the
published version that are already found in the proofs. For a
reconstruction of Hilbert's part in this episode, it seems worthwhile
to point out these characteristics here as well.

For one, Hilbert did not introduce any changes in the concluding
remarks of his paper, which distinctly echo his programmatic concern
of the 1900 Paris problems. Summarizing the main points of his note,
Hilbert concluded that
\begin{quote}
  %as all in all 
  the possibility draws near that physics will become, in
  principle, a science of the same kind as geometry: certainly the most
  magnificent glory of the axiomatic method, which here, as we see,
  takes into its service the most powerful devices of analysis, i.e.,
  the calculus of variations and invariant theory.%
  \footnote{%wie denn \"uberhaupt [...] 
    ``[...] die M\"oglichkeit naher\"uckt, da{\ss} aus der Physik im
    Prinzip eine Wissenschaft von der Art der Geometrie werde:
    gewi{\ss} der herrlichste Ruhm der axiomatischen Methode, die hier
    wie wir sehen die m\"achtigsten Instrumente der Analysis,
    n\"amlich die Variationsrechnung und Invariantentheorie, in ihre
    Dienste nimmt.'' Proofs, p.~13,
    \cite[p.~407]{Hil1915Grundlagen1}.}
\end{quote}
The emphasis of this concluding passage suggests that Hilbert saw his
communication as a culmination of his work in physics along his
program of an axiomatic treatment of physics. Using advanced and
sophisticated mathematical techniques, his aim was to promote
theoretical physics to that mathematized and axiomatized state that
geometry had already reached. Hilbert's note, as he claimed, represented
a major achievement in this endeavour.

\subsection{The gravitational field equations}

Hilbert's emphasis on the role of invariant theory and the calculus of
variations points to a feature of his theory which distinguishes it
from Einstein's \emph{Entwurf} theory and which is also already found in
the proofs.  In the \emph{Entwurf} theory, as expounded in
\cite{EinsteinA1914Grundlage}, Einstein gave a concise and independent
introduction to the mathematics of the ``absolute differential
calculus'' of Ricci and Levi-Civita in a special section headed ``From
the Theory of Covariants.''%
\footnote{``Aus der Theorie der Kovarianten.''
  \cite[pp.~1034--1054]{EinsteinA1914Grundlage}. For a discussion of
  the historical background and context of Einstein's mathematics, see
  \cite{ReichK1994Entwicklung}.} But while the concepts of this
section do provide a framework for the formulation of a theory which
is generally covariant, i.e., covariant with respect to arbitrary
coordinate transformations, gravitational field equations which
manifestly are not generally covariant were derived from a variational
principle in section D of that paper by introducing a restriction of
the theory to so-called adapted coordinate systems.  In his first
November communication Einstein had then replaced these field
equations by a different set of field equations, also derived from a
variational principle but still not generally covariant. The field
equations of his second November communication, equating the Ricci
tensor to the energy-momentum tensor, then finally were generally
covariant but these field equations were not derived from a
variational principle.

Hilbert, in his communication, introduced gravitational field equations
which are derived from a variational principle and which are generally
covariant. Thus, in contrast to Einstein's \emph{Entwurf} theory and in
contrast to Einstein's first November communication, he did not write
down gravitational field equations of restricted covariance, and, in
contrast to Einstein's second November communication, Hilbert did
formulate the generally covariant field equations in terms of a
variational principle. 

In fact, Hilbert based his theory on two ``axioms.'' The first axiom
introduces an action integral
\begin{equation}
  \int H \sqrt{g}d\tau,
\label{action}
\end{equation}
where $g$ is the determinant of the metric and $d\tau=dw_1dw_2dw_3dw_4$
with ``world parameters'' $w_i$ uniquely determining the
``worldpoints,'' i.e., the coordinates of a four-dimensional
manifold.%
\footnote{\label{note:notation}In the discussion of Hilbert's note, I
  will closely follow Hilbert's own notation. In particular,
  coordinates will be written with a subscript index, no summation
  convention is implied, and coordinate derivatives are indicated by
  simple subscript indices. A modern reader should also be aware of
  the fact that, although Hilbert does distinguish between
  contravariant, superscript and covariant, subscript indices (except
  for the coordinates), he does not raise or lower indices by means of
  the metric.}  The world function $H$ is postulated to be a function
depending on the components of the metric tensor, its first and second
derivatives, as well as on the components of the electromagnetic
four-potential and its first derivatives. The axiom specifically
postulates that the laws of physics are determined by the condition
that the variation of the integral (\ref{action}) vanish. The second
axiom then postulates that $H$ transforms as an invariant under
arbitrary transformations of the coordinates $w_i$.%
\footnote{Proofs, p.~2, \cite[p.~396]{Hil1915Grundlagen1}.}

In the proofs, the field equations are not explicitly specified. They
are only given as Lagrangian derivatives of the undetermined but
invariant ``world function'' $H$. The gravitational field equations
are given as derivatives with respect to the metric, the
electrodynamic field equations are given as derivatives with respect
to the electromagnetic potential. Hence, the gravitational field
equations appear in the form%
\footnote{Proofs, p.~3. In \cite[p.~397]{Hil1915Grundlagen1} the
  equations were given with all terms moved to the left hand side and
  set equal to 0.}
\begin{equation}
  \frac{\partial\sqrt{g}H}{\partial g^{\mu\nu}} = \sum_k
  \frac{\partial}{\partial w_k}\frac{\partial\sqrt{g}H}{\partial
    g^{\mu\nu}_k} - \sum_{k,l} \frac{\partial^2}{\partial w_k\partial
    w_l} \frac{\partial\sqrt{g}H}{\partial g^{\mu\nu}_{kl}},
\label{eq:fieldequations}  
\end{equation}
where $\mu,\nu=1, \dots 4$, $g^{\mu\nu}_k \equiv \partial
g^{\mu\nu}/\partial w_k$, $g^{\mu\nu}_{kl} \equiv \partial^2
g^{\mu\nu}/\partial w_k\partial w_l$ and factors of $1/2$ resp.\ $1/4$
are implicitly understood accounting for the symmetries in $\mu\nu$
and $kl$.  Unfortunately, the part where the world function $H$ in all
probability was specified was later cut out of the proofs. In the
published version it is given as the sum of a gravitational part $K$,
depending only on the metric and its first and second derivatives,
plus an electromagnetic part $L$, depending on the
electromagnetic potential, its derivatives, and on the metric.%
\footnote{\cite[p.~402]{Hil1915Grundlagen1}.}  In all probability,
also the excised piece of the proofs contained at least the
specification of $H$ as
\begin{equation}
  \label{eq:HKL}
  H=K+L
\end{equation}
with $K=K(g^{\mu\nu},
g^{\mu\nu}_k, g^{\mu\nu}_{kl})$ and $L=L(g^{\mu\nu}, q_s, q_{sl})$
where $q_{sl}=\partial q_s/\partial w_l$.%
\footnote{The quantities $K$ and $L$ are referred to immediately after
  the missing passage of [p.~8] of the proofs, and the splitting into
  $K+L$ is referred to later in the paper by an equation number which
  is missing exactly at that place. \label{note:KL} One possible
  reason for Hilbert's cutting out this piece would be that he wanted
  to paste it into some other manuscript in order to be spared the
  pains of copying the equations by hand. He had done a similar thing
  before in his notes for a lecture course on Euclidean geometry from
  1898/99, when he pasted some of his axioms of connection (``Axiome
  der Verbindung'') taken from \cite{Hil1895Linie} into his manuscript
  notes, see SUB Cod.\ Ms.\ Hilbert 551, pp.~10f. Another such example
  is SUB Cod.\ Ms.\ Hilbert 549 (Notes for a lecture course on
  ``Zahlbegriff und Quadratur des Kreises,'' winter 1897/98), pp.~38f.
  If the excised piece of the proofs had corresponded to the lines
  printed on the bottom of p.~402 of the published version, it would
  have contained the definitions of the Riemann curvature scalar and
  of the Ricci tensor as in eqs.~(\ref{eq:Riemannscalar}) and
  (\ref{eq:Riccitensor}).} In any case, already in the proofs, Hilbert
in a footnote insists that his world
function \emph{is} invariant whereas Einstein's is not.%
\footnote{Proofs, p.~2, \cite[p.~396]{Hil1915Grundlagen1}.}  Indeed,
the Lagrangian functions presented by Einstein either in his second
paper with Grossmann on the covariance properties of the \emph{Entwurf}
theory of spring 1914,%
\footnote{\cite[p.~219]{EinsteinAEtAl1914Kovarianzeigenschaften}.} and
in his major account of the \emph{Formale Grundlage} of General
Relativity of fall 1914,%
\footnote{\cite[p.~1076]{EinsteinA1914Grundlage}.} and in his first November
communication of 1915%
\footnote{\cite[p.~784]{EinsteinA1915Relativitaetstheorie}.}  are not
invariant. Since Hilbert's world function \emph{was} invariant and
since Lagrangian differentiation with respect to the metric is a
covariant operation, Hilbert's gravitational field equations already
in the proofs \emph{were} in any case generally covariant. And if the
excised piece of the
proofs contained the same specification as the published version,%
\footnote{\cite[p.~402]{Hil1915Grundlagen1}.}  $K$ would have been
specified as the Riemann curvature scalar,
\begin{equation}
  \label{eq:Riemannscalar}
  K=\sum_{\mu\nu}g^{\mu\nu}K_{\mu\nu},
\end{equation}
where
\begin{align}
    \label{eq:Riccitensor}
    K_{\mu\nu} = \sum_{\kappa}\left(\frac{\partial}{\partial w_{\nu}}
\left\{\begin{matrix} \mu\kappa \\ \kappa \end{matrix}\right\}\right. -
&\left.\frac{\partial}{\partial w_{\kappa}}
\left\{\begin{matrix} \mu\nu \\ \kappa \end{matrix}\right\}\right) +
\notag \\
&\sum_{k,\lambda}
\left(
\left\{\begin{matrix} \mu\kappa \\ \lambda \end{matrix}\right\}
\left\{\begin{matrix} \lambda\nu \\ \kappa \end{matrix}\right\}-
\left\{\begin{matrix} \mu\nu \\ \lambda \end{matrix}\right\}
\left\{\begin{matrix} \lambda\kappa \\ \kappa \end{matrix}\right\}
\right)
\end{align}
denotes the Ricci tensor and $\left\{\begin{matrix} \mu\nu \\ \kappa
  \end{matrix}\right\}$ are the Christoffel symbols of the second
kind. With these specifications, the resulting field equations would
differ from those of Einstein's second November memoir by a term
$(1/2)g_{\mu\nu}K$ to be subtracted from the Ricci tensor.
Nevertheless, even though Hilbert's field equations are not explicitly
given in the early version in terms of the Ricci tensor and its
contraction, Hilbert had probably realized that his theory in any case
implied field equations which differed from the ones of Einstein's
\emph{Entwurf} theory or from those put forward in Einstein's first
November communication. Indeed, the paper submitted to the G\"ottingen
Academy initially was titled ``Die Grund\emph{gleichungen} der
Physik.'' It was, however, soon
changed to ``Die Grund\emph{lagen} der Physik.''%
\footnote{``Grundgleichungen der Physik'' was the title of Hilbert's
  lecture to the G\"ottingen Mathematical Society of November 16. It
  was also the title under which Hilbert announced his communication
  in the letter of invitation circulated among the Academy members
  between November 15 and the meeting of November 20, and it was the
  title under which his communication was initially entered into the
  ``Journal'' of the ``Nachrichten'', cp.\ note \ref{note:Journal}.
  Since, however, the first proofs, which I assume to represent
  Hilbert's manuscript of November 19, already have the title
  ``Grundlagen der Physik,'' Hilbert must have changed the title
  probably with the manuscript which he gave to the printer, i.e.,
  before or on November 19. The change in title may have been a
  reaction to Einstein's claim in his letter of November 18 that
  Hilbert's equations appear to be equivalent to the ones he had found
  and presented to the Academy in his first two November
  communications.}

\subsection{The concept of energy}

Hence, the restricted covariance of the theory in the proofs does not
affect the gravitational field equations. But in the proofs the term
``Grundgleichungen'' does not refer to the gravitational field
equations alone. As a consequence of a mathematical theorem, to be
discussed in the next section, Hilbert saw that of the $14$ field
equations obtained by Lagrangian differentiation with respect to the
metric and the electromagnetic potential, only $14-4=10$ were mutually
independent. In the proofs, Hilbert then introduced four additional
equations --- of restricted covariance --- in order to satisfy the
principle of causality as Einstein had done in the \emph{Entwurf} and as
Hilbert initially thought necessary in order to guarantuee the
determinate character of the differential equations in Cauchy's sense:
\begin{quote}
  Thus, if we want to preserve the determinate character of the
  fundamental equations of physics according to Cauchy's theory of
  differential equations, we must postulate four additional
  non-invariant equations which supplement [the gravitational and
  electrodynamic field equations]%
  \footnote{``[...] so ist, wofern wir der Cauchyschen Theorie der
    Differentialgleichungen entsprechend den Grundgleichungen der
    Physik den Charakter der Bestimmtheit wahren wollen, die Forderung
    von vier weiteren zu [den Gravitations- und elektrodynamischen
    Feldgleichungen] hinzutretenden nicht invarianten Gleichungen
    unerl\"a{\ss}lich.'' Proofs, p.~4. For a historical discussion of
    the early history of the Cauchy problem in General Relativity and
    Hilbert's role in it, see \cite{StachelJ1992CauchyProblem}.}
\end{quote}
Already in his letter to Einstein of November 14, Hilbert had talked about
``the still missing 4 space-time equations.''%
\footnote{Hilbert to Einstein, 13 November 1915, CPAE8, Doc.~140.}  

In order to arrive at such additional equations, Hilbert proceeded to
define a concept of energy. Hilbert is not very explicit about the
justification for his energy concept but what he does is quite clear.
In his definition of the concept of energy he brings together
techniques from invariant theory and the calculus of variations.
He introduced what he called the ``polarisation''%
\footnote{Proofs, p.~4. See
  \cite[\S~2.]{KerschensteinerG1887Vorlesungen}, for a discussion of
  the concept of polarisation in the theory of invariants. The letter
  from Emmy Noether to Fischer, quoted above (see note
  \ref{note:Kerschensteiner}), indeed, suggests that
  \cite{KerschensteinerG1887Vorlesungen} was a standard reference on
  invariant theory in G\"ottingen at the time.} of the invariant
worldfunction $H$ with respect to an ``arbitrary contragredient
tensor'' $h^{\mu\nu}$
\begin{equation}
  J^{(h)} = \sum_{\mu,\nu}\frac{\partial H}{\partial
  g^{\mu\nu}}h^{\mu\nu} + \sum_{\mu,\nu,k}\frac{\partial H}{\partial
  g^{\mu\nu}_k}h^{\mu\nu}_k + \sum_{\mu,\nu,k,l}\frac{\partial H}{\partial
  g^{\mu\nu}_{kl}}h^{\mu\nu}_{kl}, 
\end{equation}
where the subscripts $k$ and $l$ to $h^{\mu\nu}$ denote partial
derivatives with respect to $w_k$ resp.\ $w_l$. Since polarisation is
an invariant process, Hilbert argues, $J^{(h)}$ is an invariant.
Hilbert now removes the partial derivatives on $h^{\mu\nu}$ by
treating the expression $\sqrt{g}J^{(h)}$ ``in the same manner as one
treats in the calculus of variations the integrand of a variational
problem if one wants to apply partial integration,''%
\footnote{Proofs, pp.~4, 5.} thus obtaining 
\begin{equation}
  \sqrt{g}J^{(h)} = - \sum_{\mu\nu}H\frac{\partial \sqrt{g}}{\partial
    g^{\mu\nu}}h^{\mu\nu} +
  \sum_{\mu\nu}\left[\sqrt{g}H\right]_{\mu\nu}h^{\mu\nu} + D^{(h)},
\label{eq:J_h}
\end{equation}
where $\left[\sqrt{g}H\right]_{\mu\nu}=0$ is an abbreviation for the
field equation (\ref{eq:fieldequations}), and $D^{(h)}$ a pure divergence
term. The first term, obviously, arises from the fact that
``polarisation'' is applied to the invariant $H$, whereas the whole
integrand $\sqrt{g}H$ must be varied. Without explicitly inserting the
expression into the action integral, Hilbert thus effectively derived
the gravitational field equations by variation of the action integral
with respect to the metric, if only $h^{\mu\nu}$ be regarded as an
arbitrary variation of the metric tensor. But at this point, he is
content with a reformulation of the ``polarized'' integrand in terms
of the field equations and a pure divergence term.

At this stage, Hilbert introduced the ``symmetric contravariant tensor''
\begin{equation}
\label{eq:pmn}
  p^{\mu\nu} = \sum_s\left(g^{\mu\nu}_s p^s - g^{\mu s}p^{\nu}_s -
  g^{\nu s}p^{\mu}_s\right), \quad \left(p^j_s=\frac{\partial
  p^j}{\partial w_s}\right)
\end{equation}
formed from an arbitrary vector $p^{\mu}$.%
\footnote{Note that $p^{\mu}_s$ is not obtained by lowering one index
  of $p^{\mu\nu}$, cp.\ note \ref{note:notation}.} In modern notation,
the expression is readily identified as the Lie derivative of the
metric tensor field $\underset{p}{\cal L}g^{\mu\nu} = -p^{(\mu;\nu)}$
along the vector field $p^{\mu}$ and thus represents a variation of
the metric $\delta g^{\mu\nu}$ which arises from a change of
coordinates $x^{\mu}\rightarrow x^{\mu}+p^{\mu}(x)$.%
\footnote{Hilbert only introduced $p^{\mu\nu}$ as a symmetric
  contravariant tensor without mentioning that it represents a Lie
  derivative or a variation of the metric. However, Klein, in a first
  note from January 1918 on Hilbert's first communication, refers in
  this context to Lie's ``numerous relevant publications''
  (``zahlreiche einschl\"agige Ver\"offentlichungen,''
  \cite[p.~471]{KleinF1917Note}). In a second note on Hilbert's paper
  from July 1918, Klein nevertheless found it necessary to elaborate
  on the meaning of $p^{\mu\nu}$ by an explicit calculation in an
  extra introductory paragraph
  (\cite[p.~173]{KleinF1918Differentialgesetze}).}  Hilbert now
substituted $p^{\mu\nu}$ for $h^{\mu\nu}$, and again removed all
partial derivatives on $p^s$, except for first derivatives $p^s_k$.
He thus arrived at
\begin{equation}
  \sqrt{g}J^{(p)} = - \sum_{\mu\nu}H\frac{\partial \sqrt{g}}{\partial
    g^{\mu\nu}}p^{\mu\nu} + E + D^{(p)},
\label{eq:J_p}
\end{equation}
where $D^{(p)}$ again is a divergence term and $E$ was given
explicitly as%
\footnote{It is at this point that the proofs which otherwise are of a
  fairly finished character display some irregularities which may
  indicate a certain preliminary nature of the consideration. First,
  the sentence introducing this expression is grammatically
  incomplete.  Second, Hilbert corrected the factor $p^s$ multiplying
  the first term from a misprinted $p_s$ and put an exclamation mark
  in the margin (proofs, p.~5). Third, in the remainder of the proofs,
  Hilbert mistakenly referred to his equation (10), which is our
  (\ref{eq:J_p}), where clearly his equation (9), which is our
  (\ref{eq:Eform}), was meant.}
\begin{align}
  E = &\sum\left(H\frac{\partial\sqrt{g}}{\partial
      g^{\mu\nu}}g^{\mu\nu}_s + \sqrt{g}\frac{\partial H}{\partial
      g^{\mu\nu}}g^{\mu\nu}_s + \sqrt{g}\frac{\partial H}{\partial
      g^{\mu\nu}_k}g^{\mu\nu}_{sk} + \sqrt{g}\frac{\partial
      H}{\partial g^{\mu\nu}_{kl}}g^{\mu\nu}_{skl}
  \right)p^s \notag\\
  &-\sum\left(g^{\mu s}p^{\nu}_s + g^{\nu s}p^{\mu}_s\right)
  \left[\sqrt{g}H\right]_{\mu\nu} \notag\\
  &+ \sum\left( \frac{\partial \sqrt{g}H}{\partial
      g^{\mu\nu}_{k}}g^{\mu\nu}_{s} + \frac{\partial
      \sqrt{g}H}{\partial g^{\mu\nu}_{kl}}g^{\mu\nu}_{sl} -
    g^{\mu\nu}_{s}\frac{\partial}{\partial w_l} \frac{\partial
      \sqrt{g}H}{\partial g^{\mu\nu}_{kl}}\right)p^s_k.
\label{eq:Eform}
\end{align}
Hilbert now defined the expressions $e_s$ and $e^l_s$ 
by rewriting $E$ as
\begin{equation}
  E = \sum_s e_sp^s + \sum_{s,l}e^l_sp^s_l.
\label{eq:Ee}
\end{equation}
Both $e_s$ and $e^l_s$ are well-defined by eq.~(\ref{eq:Eform}), and,
as Hilbert noted, $e_s$ is given as a total derivative with respect to
$w_s$,
\begin{equation}
  \label{eq:e_s}
  e_s = \frac{d^{(g)}\sqrt{g}H}{dw_s},
\end{equation}
where the superscript $(g)$ indicates that the electromagnetic
potentials are to be ignored in forming the derivative.

Hilbert called the expression $E$ the energy form (``Energieform''),
justifying this name by pointing out two properties. First, he noted
that a comparison of $\sqrt{g}J^{(h)}$ as given in eq.~(\ref{eq:J_h}),
where $p^{\mu\nu}$ is substituted for $h^{\mu\nu}$, with
$\sqrt{g}J^{(p)}$ as given in eq.~(\ref{eq:J_p}) shows that $E$ may be
expressed as a sum of derivatives with respect to $w_s$,
\begin{equation}
  \label{eq:Ediv}
  E = \left(D^{(h)}\right)_{h=p}-D^{(p)},
\end{equation}
if the gravitational field equations hold and hence becomes a
divergence.%
\footnote{``erh\"alt Divergenzcharakter,'' proofs, p.~6.}

The discussion of the second property, unfortunately, was mutilated by
cutting away part of the sheet on which it was given. But the argument
can nevertheless be reconstructed. Looking at $E$ as given in
(\ref{eq:Ee}) Hilbert noted that if one were to go one step further
and also remove the derivative with respect to $w_l$ in $p^s_l$ the
divergence equation ``corresponding to the energy theorem in the old
theory'',
\begin{equation}
  \label{eq:olddiv}
  \sum_l \frac{\partial e^l_s}{\partial w_l} = 0
\end{equation}
can only hold if and only if the four quantities $e_s$ vanish, i.e., if
and only if
\begin{equation}
  \label{eq:es}
  e_s = 0.
\end{equation}
The latter two equations are not generally covariant. The validity of
eq.~(\ref{eq:olddiv}) and hence the validity of eq.~(\ref{eq:es}) was
stipulated by a third axiom, and the four equations (\ref{eq:es})
complemented the ten gravitational field equations
(\ref{eq:fieldequations}) to a system of 14 equations for the 14
potentials $g^{\mu\nu}$, $q_s$.  It is these 14 equations,
(\ref{eq:fieldequations}) and (\ref{eq:es}), which Hilbert in the
proofs called the ``system of fundamental equations of
physics.''%
\footnote{Proofs, [p.~7].}

This part of the proofs, i.e., the derivation of the energy concept
was rewritten for the published version. The third axiom, introduced
in the proofs, postulated a restriction of the covariance of the
theory.  Its stipulation introduced a distinction between space-time
coordinates and world parameters, the former being those coordinates
for which the non-generally covariant energy equations
(\ref{eq:olddiv}) hold. This third axiom was later dropped in the
published version.

\subsection{Theorems about the invariant variational problem}

Both the proofs and the published paper present three mathematical
theorems, all three of them drawing consequences from the invariance of
a Lagrangian function and the action integral with respect to
arbitrary coordinate transformations.

The first theorem is a special case of a general theorem about
invariant variational problems proven three years later
by Emmy Noether and known as Noether's (second) theorem.%
\footnote{\cite[p.~239]{NoetherE1918Variationsprobleme}.} In Hilbert's
communication it is formulated for an invariant ``world function'' $J$
of the four ``world parameters'' which depends on $n$ variables and
their derivatives. The theorem then asserts that if the $n$ Lagrangian
variational equations are formed, four mutually independent linear
combinations of these $n$ differential equations and their total
derivatives are always satisfied, i.e., vanish identically. So far, the
theorem is a special case of Noether's theorem. But Hilbert went one
step further by asserting that the theorem implied that four of the
$n$ differential equations are always a consequence of the remaining
$n-4$ equations ``in the sense'' that those linear combinations hold.
The proof of this theorem was given neither in the proofs nor in the
published version but announced for another occasion.%
\footnote{Proofs, pp.~2f, \cite[pp.~397f]{Hil1915Grundlagen1}. For a
  discussion of this theorem in Hilbert's note, see
  \cite[pp.~54--69]{VizginV1994Theories}. Vizgin does not note,
  however, that Hilbert's formulation of the theorem actually goes
  beyond Noether's theorem, cp.\ also footnote \ref{note:JScomment}
  below.}

While this first theorem does not give an explicit construction of
these identities and, in fact, only asserts their existence, the other
two mathematical theorems of Hilbert's first note are formulated more
explicitly. Hilbert did not prove these theorems in the proof sheets
either and introduced their formulation by simply saying he would 
make use of them,%
\footnote{``Des Weiteren benutzen wir zwei mathematische Theoreme,
  ...'' proofs, [p.~8].} thus suggesting that they were either well-known
or trivial. In the published version, however, he did prove these
two other theorems and introduced their formulation by saying he would
now ``establish'' those two theorems.%
\footnote{``Des Weiteren stellen wir zwei mathematische Theoreme auf,
  ...'' \cite[p.~398]{Hil1915Grundlagen1}.} 

For the second theorem, Hilbert added an alternative formulation in
his published version which makes it more lucid. In this alternative
formulation, the theorem asserts that, given an invariant Lagrangian
$J$ depending on the metric components $g^{\mu\nu}$, its first and
second derivatives, and on the electromagnetic potentials $q_s$ and
its first derivatives, the following identity
\begin{equation}
    \label{eq:theorem1}
  \sum_s \frac{\partial J}{\partial w_s}p^s = PJ,
\end{equation}
holds,%
\footnote{ibid.}
where
\begin{equation}
  \label{eq:P}
  P = P_g  + P_q,
\end{equation}
\begin{align}
  \label{eq:Pg}
P_g  &= \sum_{\mu,\nu,l,k}\left( p^{\mu\nu}\frac{\partial}{\partial
      g^{\mu\nu}} + p^{\mu\nu}_{l}\frac{\partial}{\partial
      g^{\mu\nu}_{l}} + p^{\mu\nu}_{lk}\frac{\partial}{\partial
      g^{\mu\nu}_{lk}}\right), \\
  \label{eq:Pq}
P_q  &= \sum_{l,k}\left( p_l\frac{\partial}{\partial
      q_{l}} + p_{lk}\frac{\partial}{\partial q_{lk}} \right),
\end{align}
and
\begin{equation}
  \label{eq:p_l}
  p_l = \sum_s(q_{ls}p^s+q_sp^s_l),
\end{equation}
and $p_{lk}=\partial p_l/\partial w_k$.  In eq.~(\ref{eq:theorem1})
the derivative with respect to $w_s$ is, in fact, to be understood as
a total derivative, taking into account that $J$ depends on
$g^{\mu\nu}$, $g^{\mu\nu}_{l}$, $g^{\mu\nu}_{lk}$, $q_l$, and $q_{lk}$
but not explicitly on the ``world parameters'' $w_s$. In the proofs,
the theorem is only stated in a form which is obtained if the terms
implied by the left hand side of (\ref{eq:theorem1}) are subtracted
from the right hand side and the remaining terms set equal to 0.%
\footnote{Proofs, [p.~8].}

The third theorem is effectively a derivation of generalized
contracted Bianchi identities for a Lagrangian $J$ depending only on
the metric components and its first and second derivatives.  Using the
definition (\ref{eq:pmn}) of $p^{\mu\nu}$ and introducing new
notation,
\begin{equation}
  \label{eq:xxx}
  \sum_{\mu\nu}\left[\sqrt{g}J\right]_{\mu\nu}p^{\mu\nu} =
  \sum_{s,l}(i_sp^s+i_s^lp^s_l),
\end{equation}
with
\begin{align}
\label{eq:i_s}
i_s &= \quad \sum_{\mu\nu}\left[\sqrt{g}J\right]_{\mu\nu}g^{\mu\nu}_s, \\
\label{eq:i_s_l}
i_s^l &= -2\sum_{\mu}\left[\sqrt{g}J\right]_{\mu s}g^{\mu l},
\end{align}
the theorem asserts that the equation
\begin{equation}
  \label{eq:theorem3}
  i_s = \sum_l\frac{\partial i^l_s}{\partial w_l}
\end{equation}
``holds identically for all arguments, i.e., the $g^{\mu\nu}$ and their
derivatives.''%
\footnote{Proofs, p.~9, \cite[p.~399]{Hil1915Grundlagen1}. If we
  take $J$ to be the Riemann curvature scalar $K=\sum
  g^{\mu\nu}K_{\mu\nu}$, as in eqs.~(\ref{eq:Riemannscalar}),
  (\ref{eq:Riccitensor}) above, then
  $\sum_{\mu\nu}\left[\sqrt{g}J\right]_{\mu\nu}$ is equivalent to the
  tensor $\sqrt{g}G_{\mu\nu}=\sqrt{g}(K_{\mu\nu}-(1/2)g_{\mu\nu}K)$,
  and eq.~(\ref{eq:theorem3}) expresses the vanishing of its covariant
  divergence, $G^{l}_{s;l}=0$. See also
  \cite[p.~472]{KleinF1917Note}.}

\subsection{Generalized electrodynamics and Mie's theory}

Using his theorems II and III, Hilbert pointed out two features of his
general variational framework which concern the relation of the theory
to special relativistic electrodynamics and, in particular, to Mie's
theory as given in Born's interpretation. In order to appreciate
Hilbert's arguments, it is useful to briefly summarize the main points
of
Born's reinterpretation of Mie's theory.%
\footnote{\cite{BornM1914Impulsenergiesatz}.} Mie had formulated his
theory in terms of a variational principle, introducing a
Lorentz-covariant Hamiltonian depending on the fields and the
electromagnetic four-potential. Born had generalized this ansatz by
postulating a general Lorentz-covariant Lagrangian $\Phi$
depending on dynamic variables $u_{\alpha}=u_{\alpha}(x_1, x_2, x_3,
x_4)$ and their derivatives $a_{\alpha\beta}\equiv\partial
u_{\alpha}/\partial x_{\beta}$ in a four-dimensional Minkowski
spacetime with coordinates $x_{\alpha}$.  The dynamics of the general
theory was governed by the variational principle
\begin{equation}
  \label{eq:Mieprinciple}
  \delta\int \Phi(a_{ij}; u_i)dx_1dx_2dx_3dx_4 = 0,
\end{equation}
which yielded general Euler-Lagrange equations
\begin{equation}
  \label{eq:MieEulerLagrange}
  \sum_{\gamma}\frac{\partial X_{\beta\gamma}}{\partial x_{\gamma}} -
  X_{\beta} = 0,
\end{equation}
where
\begin{equation}
  \label{eq:X}
  \frac{\partial\Phi}{\partial a_{\alpha\beta}} = X_{\alpha\beta},
\qquad
  \frac{\partial\Phi}{\partial u_{\alpha}} = X_{\alpha}.
\end{equation}
Interpreting the dynamical variables $u_{\alpha}$ as displacements of a
four-dimensional continuum, the ansatz could be specialized to special
relativistic elasticity theory. In this case $\Phi$ would depend only
on 6 combinations $e_{ij}$ ($i,j=1,2,3$) of the $a_{\alpha\beta}$,
representing the deformation quantities introduced by Born in his
definition of relativistic rigidity.%
\footnote{\cite[pp.~10,~15]{BornM1909Theorie}, \cite{HerglotzG1911Mechanik}.}
Alternatively, the dynamic variables $u_{\alpha}$ may be regarded as
the components of the electromagnetic four-potential. This would give
the specialization corresponding to Mie's theory. In this case $\Phi$
only depended on the antisymmetric combinations
$a_{\alpha\beta}-a_{\beta\alpha}$, representing the components of the
electromagnetic field tensor. The main point of Born's paper was the
identification of Mie's energy-momentum tensor in this generalized
framework. In the general Lagrangian formulation, the Euler-Lagrange
equations (\ref{eq:MieEulerLagrange}) entail conservation laws of
the form
\begin{equation}
  \label{eq:BornConservation}
  \frac{\partial\Phi}{\partial x_{\alpha}} =
  \sum_{\gamma}\frac{\partial}{\partial x_{\gamma}}(\sum_{\beta}
  X_{\beta\gamma}a_{\beta\alpha}),
\end{equation}
if $\Phi$ does not depend explicitly on the coordinate
$x_{\alpha}$.%
\footnote{This assumption distinguishes Mie's theory from the usual
  Maxwell theory with charges and currents as external sources as
  given by the usual Lorentz electron theory. This theory can formally
  be included into the general framework by letting $\Phi$ depend on
  the external sources, however, then $\Phi$ would explicitly depend on the
  space-time variables.}  These conservation laws then justify the
definition of the energy momentum tensor in terms of the Lagrangian
$\Phi$ as
\begin{equation}
  \label{eq:T}
  T_{\alpha\beta} =
  \Phi\delta_{\alpha\beta}-\sum_{\gamma}a_{\gamma\alpha}X_{\gamma\beta},
\end{equation}
(now known as the ``canonical'' energy-momentum tensor)
since it could then be written in the compact notation of four-dimensional
vector calculus as
\begin{equation}
  \label{eq:Div}
   \operatorname{Div} T = 0.
\end{equation}
The point of Born's paper, discussed in its final paragraph, was to
show that the energy-momentum tensor of Mie's theory is not simply
given by eq.~(\ref{eq:T}) since it has to depend only on the
antisymmetric combinations $a_{\gamma\alpha}-a_{\alpha\gamma}$. Rather
it is obtained by adding a tensor
$a_{\alpha\gamma}X_{\gamma\beta}-u_{\alpha}X_{\beta}$ with vanishing
four-divergence such that it is the (symmetric)%
\footnote{\cite[p.~533]{MieG1912Grundlagen1},
  \cite[p.~35]{BornM1914Impulsenergiesatz}.}
tensor
\begin{equation}
  \label{eq:TMie}
  S_{\alpha\beta} =
  \Phi\delta_{\alpha\beta}-\sum_{\gamma}(a_{\gamma\alpha}-a_{\alpha\gamma})X_{\gamma\beta} - u_{\alpha}X_{\beta}
\end{equation}
that represents the energy-momentum of Mie's theory.

It is to this reformulation of Mie's electrodynamics by Born that Hilbert
referred to in the following part of his paper. By his first axiom,
Hilbert emphasized, in the proofs as well as in the published version,
that his world function $H$ would not only depend on the metric tensor
and its first and second derivatives but also on the electromagnetic
potentials $q_s$ and its first derivatives $q_{sl}$. In fact, Hilbert
emphasized in a footnote that while the general use of a variational
principle was a characteristic of Mie's theory, the particular use
of the arguments $q_s$ and $q_{sk}$ was introduced by Born.%
\footnote{Proofs, p.~2, \cite[p.~396]{Hil1915Grundlagen1}.}

Hilbert now pointed out two consequences following from the assumption
of general invariance for $H$. From his Theorem II, Hilbert deduced
the following relation for the electromagnetic part $L$ of the world
function
\begin{align}
  \sum_{\mu,\nu,m}\frac{\partial L}{\partial g^{\mu\nu}}(g^{\mu
    m}p^{\nu}_m &+ g^{\nu m}p^{\mu}_m)
  -\sum_{s,m}\frac{\partial L}{\partial q_s} q_mp^m_s \notag \\
  &- \sum_{s,k,m}\frac{\partial L}{\partial
    q_{sk}}(q_{sm}p^m_k+q_{mk}p^m_s+q_mp^m_{sk})=0,
  \label{eq:(21)}
\end{align}
since, by assumption, $L$ does not depend on the derivatives of the metric.
By setting the coefficient in front of $p^m_{sk}$ equal to 0, Hilbert
then deduced that $L$ could, in fact, only depend on the antisymmetrized
derivatives
\begin{equation}
  \label{eq:M}
  M_{ks}=q_{sk}-q_{ks}.
\end{equation}
Thus, Hilbert had shown that it followed from his second axiom, i.e.,
from the postulate of general covariance, that the world function
$H=K+L$ can indeed only depend on the antisymmetrized derivatives of
the electromagnetic potential, i.e., on the field tensor, although the
electromagnetic part $L$ of world function $H$ initially was
introduced as a function depending on all derivatives of $q_s$.%
\footnote{Proofs, p.~10, \cite[p.~403]{Hil1915Grundlagen1}.}
Hence Hilbert had shown that, in the generalized framework, this feature
of Maxwellian electromagnetic theory was a consequence of the
invariance of the world function with respect to arbitrary coordinate
transformations:
\begin{quote}
This result, which first gives the special character of the Maxwell
equations, arises here essentially as a consequence of the general
invariance, i.e., on the basis of axiom II.%
\footnote{``Dieses Resultat, durch welches erst der Charakter der
  Maxwellschen Gleichungen bedingt ist, ergibt sich hier wesentlich
  als Folge der allgemeinen Invarianz, also auf Grund von Axiom II.''
  \cite[p.~403]{Hil1915Grundlagen1}. In the proofs, p.~10, the
  explanatory half-sentence ``durch welches ... bedingt ist'' is
  missing.}
\end{quote}

The second consequence of the postulate of general invariance directly
referred to Mie's energy-momentum tensor. Again referring to
eq.~(\ref{eq:(21)}), Hilbert noted that setting the coefficient of
$p^{\nu}_m$ equal to 0 and taking into account that $L$ depends only
on the field variables $M_{ks}$ as defined in (\ref{eq:M}) one obtains
for the electromagnetic energy the expression
\begin{equation}
  \label{eq:(24)}
  -2\sum_{\mu}\frac{\partial\sqrt{g}L}{\partial g^{\mu\nu}}g^{\mu m} =
   \sqrt{g} \left\{L\delta^m_{\nu}-\frac{\partial L}{\partial
   q_m}q_{\nu} -\sum_s\frac{\partial L}{\partial M_{ms}}M_{\nu s}\right\}.
\end{equation}
Comparing this expression with the corresponding expression (\ref{eq:TMie})
of Born's theory, Hilbert noted that Mie's energy tensor is obtained
in his generally covariant, variational framework by
generally covariant operations and subsequent specialization to the
case of special relativity,%
\footnote{Proofs, p.~10, \cite[p.~404]{Hil1915Grundlagen1}.}  an
insight obtained at least by mid-October and already then communicated
to Sommerfeld and an insight which Hilbert had taken great pleasure in
(``Hauptvergn\"ugen''), as he reports to Einstein in his
correspondence of November 13. Nowadays, the energy tensor of any
theory is obtained by variation of the matter Lagrangian with respect
to the metric, but for Hilbert this result was intimately linked with
his understanding of the theory as a generalization of Mie's
electrodynamic field theory of matter. In his paper, Hilbert
summarized this insight in the following words:
\begin{quote}
  Hence, Mie's electromagnetic energy tensor is nothing but that
  generally covariant tensor which is obtained by differentiation of
  the invariant $L$ with respect to the gravitation potentials
  $g^{\mu\nu}$ in that limiting case [of special relativity]% 
  \footnote{``[...] der Mie'sche elektromagnetische Energietensor ist
    also nichts anderes als der durch Differentiation der Invariante
    $L$ nach den Gravitationspotentialen $g^{\mu\nu}$ entstehende
    allgemeine invariante Tensor beim \"Ubergang zu jener Grenze [der
    speziellen Relativit\"atstheorie] ...'' proofs, p.~10,
    \cite[p.~404]{Hil1915Grundlagen1}.}
\end{quote}
and immediately added that this was
\begin{quote}
  --- a fact which pointed me in the first place to the necessary and
  close relation between Einstein's general theory of relativity and
  Mie's electrodynamics and which convinced me of the
  correctness of the theory developed here.%
  \footnote{``--- ein Umstand, der mich zum ersten Mal auf den
    notwendigen engen Zusammenhang zwischen der Einsteinschen
    allgemeinen Relativit\"atstheorie und der Mie'schen Elektrodynamik
    hingewiesen und mir die \"Uberzeugung von der Richtigkeit der hier
    entwickelten Theorie gegeben hat.'' proofs, p.~10,
    \cite[p.~404]{Hil1915Grundlagen1}.}
\end{quote}

\subsection{Gravitation, electromagnetism, and the theory of matter}

The first mathematical theorem of Hilbert's paper, as discussed above,
asserted the existence of four independent identities between the
fourteen field equations obtained by Lagrangian differentiation of the
action integral. This mathematical theorem was interpreted by Hilbert
as implying that the generalized Maxwell equations may be regarded as
a consequence of the gravitational field equations in the sense that
the four linear combinations of the Maxwell equations and their total
derivatives identically vanish. Thus Hilbert wrote
\begin{quote}
  As a consequence of that mathematical theorem we can immediately
  assert that, in the sense indicated, the electrodynamical phenomena
  are effects of gravitation.%
  \footnote{``[...] wir k\"onnen unmittelbar wegen jenes
    mathematischen Satzes die Behauptung aussprechen, da{\ss} in dem
    bezeichneten Sinne die elektrodynamischen Erscheinungen Wirkungen
    der Gravitation sind.''  Proofs, p.~3,
    \cite[p.~397]{Hil1915Grundlagen1}.}
\end{quote}
and, adding even more emphasis to this feature of his theory, he
called this insight into the relation between electromagnetism and
gravitation the ``Leitmotiv'' of his theory.%
\footnote{Proofs, p.~2, \cite[p.~396]{Hil1915Grundlagen1}.}
However, although Hilbert seemed to believe that the validity of the
gravitational field equations implied the validity of the generalized
Maxwell equations, he concretized his assertion by deriving the
precise mathematical form of those four identities in the final part
of his paper.

Starting from the electromagnetic energy as given in (\ref{eq:(24)})
as derivative of $\sqrt{g}L$ with respect to $g^{\mu\nu}$, Hilbert
used the gravitational field equations
\begin{equation}
  \label{eq:(26)}
  \left[\sqrt{g}K\right]_{\mu\nu} + \frac{\partial\sqrt{g}L}{\partial
      g^{\mu\nu}} = 0,
\end{equation}
which hold for the ansatz $H=K+L$,%
\footnote{Cp.\ footnote \ref{note:KL} above.} and the definition
(\ref{eq:i_s_l}) of $i^l_s$ in order to express the left hand side of
(\ref{eq:(24)}) as
\begin{equation}
  \label{eq:i_nu}
  -2\sum_{\mu}\frac{\partial\sqrt{g}L}{\partial g^{\mu\nu}}g^{\mu m} =
  -i^m_{\nu}.
\end{equation}
In order to invoke the contracted Bianchi identity
$i_{\nu}=\sum\partial i^m_{\nu}/\partial w_{m}$ of his theorem III,
Hilbert formed the coordinate divergence of the right hand side of
(\ref{eq:(24)}) and obtained
\begin{equation}
  i_{\nu} = \sum_m \frac{\partial}{\partial w_m} \left(
    -\sqrt{g}L\delta^m_{\nu} + \frac{\partial\sqrt{g}L}{\partial
      q_m}q_{\nu} + \sum_s \frac{\partial\sqrt{g}L}{\partial
      M_{sm}}M_{s\nu} \right),
\end{equation}
and by further transformation of this expression, introducing the
Lagrangian derivatives of the generalized Maxwell equations
\begin{equation}
  \label{eq:Maxwell}
  \left[\sqrt{g}L\right]_m = \frac{\partial\sqrt{g}L}{\partial q_m} -
  \sum_s\frac{\partial}{\partial w_s}\frac{\partial\sqrt{g}L}{\partial
    q_{ms}}
\end{equation}
and by making use of the antisymmetry of $M_{sm}$, the definition
(\ref{eq:i_s}) of $i_{\nu}$, and the relation $\partial L/\partial
q_{sm}=-\partial L/\partial q_{ms}$, Hilbert finally arrived at the
four identities
\begin{equation}
  \label{eq:(32)}
  \sum_m\left(M_{m\nu}\left[\sqrt{g}L\right]_m+q_{\nu}\frac{\partial}{\partial
      w_m}\left[\sqrt{g}L\right]_m\right)=0.
\end{equation}
He summarized the result by repeating his claim that
\begin{quote}
  the gravitational equations [...] entail indeed the four
  mutually independent linear combinations (\ref{eq:(32)}) of the
  electrodynamic fundamental equations [...] and their first derivatives%
  \footnote{\label{note:JScomment} ``[...] aus den
    Gravitationsgleichungen [...] folgen in der Tat die vier von
    einander unabh\"angigen linearen Kombinationen (\ref{eq:(32)}) der
    elektrodynamischen Grundgleichungen [...] und ihrer ersten
    Ableitungen.'' Proofs, p.~12, \cite[p.~406]{Hil1915Grundlagen1}.
    As John Stachel pointed out in the discussion at the G\"ottingen
    workshop, the relations (\ref{eq:(32)}) do not imply the
    electromagnetic field equations themselves. From the second term
    it follows that for $q_{\nu}\neq 0$ and given field equations
    $\left[\sqrt{g}L\right]_m=0$ on an initial hypersurface
    $w_4=\textrm{const}$, one has
    $\partial\left[\sqrt{g}L\right]_4/\partial w_4=0$, but no
    restrictions for $\partial\left[\sqrt{g}L\right]_i/\partial w_4$,
    $i=1$, $2$, $3$ are implied. Therefore, even if the
    electromagnetic field equations hold on an initial hypersurface,
    they will not continue to hold automatically off it as a
    consequence of the identities.  The second term in
    eq.~(\ref{eq:(32)}) vanishes if $L$ is gauge invariant. In this
    case the electromagnetic field equations do follow algebraically
    if $M_{m\nu}$ has an inverse. The relation was derived in this
    form in \cite[p.~473]{KleinF1917Note} who had started from an
    electromagnetic part $L=\alpha
    M_{\nu\mu}M_{\sigma\rho}(g^{\mu\rho}g^{\nu\sigma}-g^{\mu\sigma}g^{\nu\rho})$.
    Therefore, Klein can agree with Hilbert's claim that the Maxwell
    equations follow from the gravitational field equations.}
\end{quote}
and emphasized:
\begin{quote}
  This is the full mathematical expression of the above asserted
  proposition about the character of electrodynamics as a
  phenomenon following from gravitation.%
  \footnote{``Dies ist der ganze mathematische Ausdruck der oben
    allgemein ausgesprochenen Behauptung \"uber den Charakter der
    Elektrodynamik als einer Folgeerscheinung der Gravitation.''
    ibid. \cite[p.~406]{Hil1915Grundlagen1} changed ``ganze'' to
    ``genaue.''}
\end{quote}

Finally, the freedom to specify the world function by non-linear so
long as invariant terms of the electromagnetic potential in the sense
of Mie opened up the possibility that non-linear generalizations of
Maxwell's equations may be obtained, the solutions of which would
properly describe the electron. It was the computation of the electron
in this sense that Hilbert hoped to achieve before he wrote down his
theory for print, and which he now left for the second part of his
paper.%
\begin{quote}
  Further elaboration and, in particular, the 
  application of my basic equations to the fundamental questions
  of electricity theory I defer to a later communication.%
  \footnote{``Die genauere Ausf\"uhrung sowie vor Allem die spezielle
    Anwendung meiner Grundgleichungen auf die fundamentalen Fragen der
    Elektrizit\"atslehre behalte ich sp\"ateren Mitteilungen vor.''
    Proofs, p.~1, \cite[p.~395]{Hil1915Grundlagen1}. See also Hecke to
    Hilbert, 7 March 1916, ``When will you publish the electron?''
    (``Wann publizieren Sie das Elektron?'') SUB Cod.\ Ms.\ Hilbert
    141/7.}
\end{quote}

\section{A speculative reconstruction}

Given these characteristics of Hilbert's theory which are already
found in the first proofs, the following scenario of Hilbert's path
towards his unified field theory of electromagnetism and gravitation,
including generally covariant gravitational field equations and
generalized Maxwell equations in the sense of Mie, is suggested. The
scenario would be the following: being interested in fundamental
theories of physics, Hilbert studied Mie's theory as early as 1913 as
a novel and attractive field theory of matter. An important step was
Born's work of late 1914 on Mie's theory, which reinterpreted Mie's
variational ansatz in terms of special-relativistic continuum
mechanics, pointing out formal relations to Herglotz's
special-relativistic elasticity theory, which Hilbert knew well.
Hilbert had also heard about Einstein's and Grossmann's theory as
early as December 1913 through a talk at the G\"ottingen Mathematical
Society. Just as in the case of Mie's theory, Hilbert must have had a
genuine interest in attempts at generalizing the special theory of
relativity, for Hilbert the heritage of his friend Minkowski's
geometric reinterpretation of the Lorentz transformations.  When
Hilbert then succeeded in inviting Einstein to come to G\"ottingen he
was intrigued by the latter's presentation. As Einstein found, Hilbert and
the other G\"ottingen mathematicians were completely convinced by the theory.%
\footnote{\cite[p.~259]{PaisA1982Subtle}, CPAE8, Doc.~110, cp.\ note
  \ref{note:EHcorrespondence}.}

Thus, when Hilbert went away for his summer vacation to Switzerland
and R\"ugen, he had the following situation to ponder on.  Both Mie's
theory, in particular as reinterpreted by Born, and Einstein's theory
of 1914 started from a variational principle with a Lagrangian of a
certain covariance group. Mie's theory was invariant with respect to
the orthogonal group of special relativity. The mathematics of
Einstein's theory was set up to be invariant with respect to general
diffeomorphisms but in his 1914 exposition of the \emph{Formale
  Grundlage der allgemeinen Relativit\"atstheorie} Einstein had
expounded the hole argument justifying the necessity of restricting
the covariance of the theory and elaborated an intricate mathematical
argument for the unique specification of the correct Lagrangian in
accordance with his so-called ``adapted'' coordinate systems. The
argument was based on a rather pedestrian treatment of the covariance
properties of the action integral, distinguishing between somewhat
hybrid variations restricted by Einstein's concept of ``justified'' coordinate
transformations. Now, invariant theory and the calculus of variations
were fields of Hilbert's expertise, and when Einstein presented his
considerations in July 1915 in G\"ottingen, he found that
everything was understood, including the details.  Nevertheless,
Einstein's mathematical treatment of adapted coordinate systems must
have appeared awkward to Hilbert just as they
had appeared fishy to the mathematician Levi-Civita.%
\footnote{For a discussion of Levi-Civita's objections to Einstein's
  proof of the covariance properties of the field equations in
  \cite{EinsteinA1914Grundlage} expressed in a correspondence between
  him and Einstein in the spring of 1915, see
  \cite{CattaniCEtAl1989Controversy}. Whatever Hilbert's objections to
  Einstein's theory were, they referred in any case to Einstein's way
  of handling variations of the metric, see Einstein to Hilbert, 30
  March 1916, CPAE8, Doc.~207.}

Mie's theory on the other hand had other problems. For one, Mie's
theory of gravitation might not have been too attractive. Also, Mie had
specified his world function by adding a term of sixth power in the
potential and had explicitly integrated the resulting differential
equation.%
\footnote{\cite[pp.~18ff]{MieG1912Grundlagen2}.}  The result was
unsatisfactory to Mie because the resulting electrons would neither be
indivisible nor stable. Mie's electrons would tend to attract each
other and then simply merge into a single huge lump of charge.%
\footnote{\cite[p.~38]{MieG1912Grundlagen2}.}  However, his theory
formulated a research program by posing the problem of finding a
satisfactory Lagrangian function. The research program was formulated
by Born as follows:
\begin{quote}
  In earlier times, one formulated the goal of the mechanical
  explanation of nature, to derive all physical and chemical
  properties of matter by assuming a Lagrangian function $\Phi$ for
  the interaction of the atoms. In the same way Mie now poses the
  problem of chosing his ``world function'' $\Phi$ such that from it
  follows on the basis of his differential equations the existence of
  the electron and of the atoms, as well as all their interactions. I
  should like to see this problem as the mathematical content of that
  program which regards the establishment of an ``electrodynamic
  worldview'' as the aim of
  physics.%
  \footnote{``Wie man fr\"uher als Ziel der mechanischen
    Naturerkl\"arung hinstellte, durch Annahme einer Lagrangeschen
    Funktion $\Phi$ f\"ur die Wechselwirkung der Atome alle
    physikalischen und chemischen Eigenschaften der Materie
    abzuleiten, so stellt jetzt Mie die Aufgabe, seine `Weltfunktion'
    $\Phi$ so zu w\"ahlen, da{\ss} daraus auf Grund seiner
    Differentialgleichungen sowohl die Existenz des Elektrons und der
    Atome, als auch die Gesamtheit ihrer Wechselwirkungen hervorgehen.
    Diese Forderung Mies m\"ochte ich als den mathematischen Inhalt
    jenes Programms ansehen, das die Aufstellung eines
    `electromagnetischen Weltbildes' als Ziel der Physik hinstellt.''
    \cite[pp.~24]{BornM1914Impulsenergiesatz}.}
\end{quote}
And Hilbert, who knew well both the relation between a variational
ansatz and the implied differential equation and the complexity of
non-linear differential equations, regarded Mie's research program, as
formulated here by Born, as the real challenge of theoretical
physics.

One of Hilbert's first ideas might have been to generalize the
covariance of Mie's theory and to integrate it into a common
mathematical framework of both Mie's and Einstein's theories.
Mathematically, something like the metric tensor introduced by
Einstein would also show up if one simply allowed for general
coordinate transformations in Mie's theory. Why not let Mie's world
function depend on the metric components as well, and add it to the
Lagrangian of Einstein's theory?  Using generally covariant
operations, one might then see whether Mie's results are obtained in
the special relativistic limit, which would give concrete physical
meaning to the generalized theory. This indeed proved to be the case
for Mie's energy tensor. Would it also hold for the electron? Hilbert
might then have investigated general properties of the resulting
differential equations, drawing on his knowledge of partial
differential equations and on knowledge about what we would now call
the Lie derivative. This might have given him the insight that
covariance with respect to diffeomorphisms yielded additional
relations between the field equations, a result which immediately
suggested that the ten gravitational field equations and the Maxwell
equations were not independent.

These are vague considerations, of course. The Hilbert archive
contains a number of disparate calculations, which are hard to
identify or even only to date. There is opportunity for future research.
In any case, Hilbert obtained most of the characteristic features of
his first note already in the fall of 1915, with the exception of the
explicit form of the field equations in terms of the Ricci tensor and
the realization that the restrictive implications of energy
conservation are but a remnant dark spot of Einstein's previous
misconceptions.

\section{Aftermath}

Five days after Hilbert had submitted his communication to the
G\"ottingen Academy, Einstein presented a fourth communication to the
Berlin Academy.%
\footnote{\cite{EinsteinA1915Feldgleichungen}.} In this communication
Einstein presented his final field equations, the Einstein equations
where a trace term was added to the energy-momentum tensor. The
introduction of the trace term had two decisive consequences. First,
the requirement of energy conservation no longer implied restrictions
on the permissible coordinates. The theory now was fully covariant. A
distinction, in Hilbert's terms, between ``world parameters'' and
``space-time coordinates'' was no longer necessary.  Second, the
gravitational field equations no longer implied that the trace of the
energy-momentum tensor of matter had to vanish.  Hence, the field
equation no longer implied anything about the constitution of matter.

We don't know whether or when Einstein sent copies of his final
communication to Hilbert, but we may assume that he did, just as he
had sent his first communication to Hilbert, and in any case
Einstein's final communication was available in print by December 2.%
\footnote{CPAE6, p.~244.}  

Hilbert, therefore, had occasion to think about his own theory in the
light of Einstein's last November communication. And in any case, he
continued working on his theory. Thus, roughly a week later, on
November 30, he and Caratheodory gave a talk to the G\"ottingen
Mathematical Society on invariant theory.%
\footnote{JDM 24 (1915), p.~111.}  And four
days later, on December 4, Hilbert presented a second communication on
the ``Foundations of Physics'' to the G\"ottingen Academy. About the
content of this second communication nothing is known since its
publication was postponed, at least for the time being.%
\footnote{GM 1916, p.~6. The ``Journal'' for the
  ``Nachrichten'' (GAA Scient 66, Nr.~2, cp.\ note
  \ref{note:Journal}), entry Nr.~731, says ``wiederholt in No~739.''
  Entry Nr.~739 of the ``Journal'' lists a communication by Hilbert,
  titled ``Grundlagen der Physik Zweite Mitteilung,'' presented on 26
  February. The date of 2 March 1916 of this entry in the column ``Zum
  Druck'' was deleted and in the last column its says ``=No. 731,
  759.'' Entry Nr. 759 finally is Hilbert's second communication:
  ``Grundlagen der Physik (Zweite Mitteilung),'' presented on 23
  December 1916, sent to the printer on 29 January 1917, and published
  as \cite{Hil1917Grundlagen2}.} On that Academy meeting of December
4, Hilbert also presented a note by Emmy Noether with the title
``Kr\"ummungsinvarianten im mehrdimensionalen Raume'' for publication
in the Academy's \emph{Nachrichten}. We don't know either
what the point of this paper was since it was later officially withdrawn%
\footnote{The ``Journal'' for the ``Nachrichten'' (GAA Scient 66,
  Nr.~2, cp.\ the previous note and note \ref{note:Journal}) in this
  case says ``zur\"uckgezogen,'' see entry Nr.~733, and indeed a paper
  with this title was never published by Emmy Noether; see also
  GM 1916, p.~6 where Noether's paper was also
  announced.} and the manuscript appears to be lost.  It is known,
however, that on the same day, December 4, 1915, Hilbert wrote to the
Prussian minister on behalf of Noether's habilitation. In this letter
he wrote:
\begin{quote}
  ... also the mathematical-physical developments (Einstein's theory of
  gravitation, theory of time and space) are presently moving towards
  an unforeseen point of culmination; and in this matter Miss
  Noether is my most successful collaborator ...%
\footnote{``... auch die mathematisch-physikalischen Fortschritte
  (Gravitationstheorie von Einstein, Theorie von Zeit und Raum)
  streben einem ungeahnten Kulminationspunkt gerade gegenw\"artig zu;
  und da habe ich hier Frl. Emmy Noether als erfolgreichste
  Mitarbeiterin ...'' Hilbert to the Prussian Ministry, 4 December
  1915, quoted in \cite[p.~22]{TollmienC1991Habilitation}.}
\end{quote}

The next day, December 5, Hilbert, Klein, Voigt, Runge, and Wiechert
wrote a proposal to the G\"ottingen Academy suggesting that Einstein
be elected as a corresponding member.%
\footnote{See note \ref{note:election} below.}  Two days later, on
December 7, Hilbert and Caratheodory continued their lecture on
invariant theory in the G\"ottingen Mathematical Society.%
\footnote{JDM 24 (1915), p.~111.}

With Hilbert's communication in press and Einstein's final
communication already published, the hectic feverishness of November now
abated somewhat. It seems that Hilbert took the occasion of
Einstein's election as a corresponding member of the G\"ottingen
Academy, which took place on the Academy meeting of December 18, to
inform him about his election in a personal letter before the official notice.%
\footnote{\label{note:election}The chronology of Einstein's election
  as a corresponding member of the G\"ottingen Academy is as follows:
  on November 22, the secretary of the mathematical-physical class,
  Ehlers, had sent out an invitation to an ``unverbindliche
  Vorbesprechung'' on Saturday, November 27, following a request by
  Klein (``Nach einem Wunsche des Herrn F.~Klein'' (GAA Pers 10.3, 25).
  It was in all probability Klein who took the initiative about the
  elections to the Academy in December 1915; see notes about the
  Academy's members and a list of possible candidates in SUB Cod.\ 
  Ms.\ Klein III F, sheets~9--11, dated 29 November 1915. The proposal
  for the election of Einstein (``einer der tiefsten und
  erfolgreichsten Forscher im Gebiete der theoretischen Physik'') as a
  corresponding member
  was dated December 5 and was signed by Hilbert, Klein, Runge, Voigt,
  and Wiechert (GAA Pers 20, 913). On December 8, Ehlers invited the
  members of the class to the meeting on Saturday, December 11,
  announcing that it had been suggested to elect P.~Debye and
  H.~Stille as full (``ordentlich'') members, G.~Cantor as external
  (``ausw\"artig'') member, and A.~Einstein as well as P.~Koebe as
  corresponding (``korrespondierend'') members (GAA Pers 10.3, 25).
  The final election then took place at the joint meeting of both
  classes on December 18 (GAA Chron 2.1, Vol.~6, 413). The official
  notice was sent out on December 22 (GAA Chron 8, Vol.~6). Einstein
  acknowledged his election in a letter to the Academy dated December
  23 (GAA Pers 20, 416).
  
  In 1923, Born suggested that A.~Einstein and N.~Bohr be elected as
  external members of the Academy (GAA Pers 19, 498). Einstein's
  letter of acknowledgment of 17 October 1923 contains the following
  amusing postscript:
\begin{quote}
  ``Ich vermute, dass ich schon Mitglied der `Gesellschaft der
  Wissenschaften in G\"ottingen' bin, da ich seit einigen Jahren die
  G\"ottinger Nachrichten erhalte. Ich erlaube mir, dies beizuf\"ugen
  um zu verh\"uten, dass ich irrt\"umlich ein zweites Mal gew\"ahlt
  werde.'' (GAA Pers 19, 501)
\end{quote}
On November 28, 1933, after several inquiries about Einstein at other
academies and also at the German Embassy in Washington, the Academy's
Secretary informed the Curator of G\"ottingen University that Einstein
had been deleted from the list of external members of the Academy (GAA
Pers 66, 18--23). Letters of reconciliation were sent out to the
excluded members in August 1945 and again in December 1946 (GAA Pers
6, 10, 36). To these letters, Einstein did not respond.} It is in
response to this letter that Einstein himself wrote a conciliatory
response to Hilbert after having somewhat settled his own emotional
turmoil around his November breakthrough, writing
\begin{quote}
  It is objectively a pity if two guys who have somewhat worked their
  way free of this shabby world don't take pleasure in one another.%
  \footnote{``Es ist objektiv schade, wenn sich zwei wirkliche Kerle,
    die sich aus dieser sch\"abigen Welt etwas herausgearbeitet haben,
    nicht gegenseitig zur Freude gereichen.'' Einstein to Hilbert, 20
    December 1915, CPAE8, Doc.~167.}
\end{quote}

After having been forced to a somewhat premature write up of his
results, Hilbert now reconsidered the calculations in the first set of
proofs for his communication, in particular those about the energy
theorem. It probably took Hilbert some time but eventually he found
that he could indeed derive an energy expression which also gave Mie's
energy tensor in the limit of special relativity but which satisfied a
conservation law that was not of restricted covariance.  The latter
feature was in agreement with Einstein's fourth and final commmunication of
November 25 which contained the final field equations of Einstein's
theory. The additional trace term introduced in that communication to
the gravitational field equations implied that energy-momentum
conservation no longer posed a restriction on the admissible
coordinates. Now Hilbert also found an energy expression which,
together with his field equation, was identically conserved without
imposing additional constraints. It is not clear when Hilbert achieved
this insight but on January 25, 1916, he gave a lecture to the G\"ottingen
Mathematical Society on ``invariant theory and the general energy
theorem.''%
\footnote{JDM 25 (1917), p.~31.} And since his final
communication was probably in press by mid-February, we may assume
that he had found his new energy expression by that time and that his
talk of January 25 presented a discussion of the energy concept as
contained in the published version of his communication.

In the proofs, as discussed above, Hilbert had derived his energy
concept by looking at the expression $\sqrt{g}J^{(p)}$ where $J^{(p)}$
was the ``polarisation'' of the invariant $H$ with respect to the
metric $g^{\mu\nu}$. The result had been an ``energy form''
\begin{equation}
  E = e_sp^s + e_s^lp^s_l
\end{equation}
which could also be written in terms of the gravitational field
equations and a pure divergence term as
\begin{equation}
  E = -\left[\sqrt{g}H\right]_{\mu\nu}p^{\mu\nu} 
       + \left(D^{(h)}_{(h=p)}-D^{(p)}\right).
\end{equation}
The term $e_s$ had been identified as a total derivative of
$\sqrt{g}H$ taking into account only the dependence of the world
function $H$ on the metric and its derivatives.

In the published paper, Hilbert now formed a ``polarisation'' of
$\sqrt{g}H$ with respect to all variables, the metric components and
the electromagnetic potential. In the notation of his Theorem II, he
considered the expression $P(\sqrt{g}H) =
P_g(\sqrt{g}H)+P_q(\sqrt{g}H)$. Now Hilbert's aim was to derive an
expression which only depended on the arbitrary vector $p^s$ linearly,
and which had a vanishing coordinate divergence. He showed that
$P(\sqrt{g}H)$ could be written as
\begin{equation}
\label{eq:Penergy}
  P(\sqrt{g}H) = \sum_{\mu\nu} \left[\sqrt{g}H\right]_{\mu\nu} p^{\mu\nu} + 
      \sum_k \left[\sqrt{g}H\right]_kp_k + 
      \frac{\partial\sqrt{g}(a^l+b^l+c^l+d^l)}{\partial w_l}
\end{equation}
where
\begin{align}
\label{eq:a_l}
  a^l &= \sum_{\mu,\nu,k}\frac{\partial H}{\partial
  g^{\mu\nu}_{kl}}
\left\{p^{\mu\nu}_k + \sum_{\rho}\left(\left\{\begin{matrix} k \rho \\ \mu \end{matrix}\right\}p^{\rho\nu} +
\left\{\begin{matrix} k \rho \\ \nu \end{matrix}\right\}p^{\rho\mu}
\right)\right\},\\ \label{eq:b_l}
b^l &= \sum_{\mu,\nu,\rho,k}\left(
\frac{\partial H}{\partial g^{\mu\nu}_{l}} -
\frac{\partial}{\partial w_k}\frac{\partial H}{\partial g^{\mu\nu}_{lk}} -
\frac{\partial H}{\partial g^{\rho\nu}_{lk}}
\left\{\begin{matrix} k \mu \\ \rho \end{matrix}\right\} -
\frac{\partial H}{\partial g^{\mu\rho}_{lk}}
\left\{\begin{matrix} k \nu \\ \rho \end{matrix}\right\}
\right)p^{\mu\nu},\\ \label{eq:c_l}
c^l &= \sum_{kl}\frac{\partial H}{\partial q_{kl}}p_k,\\ \label{eq:d_l}
d^l &= \frac{1}{2\sqrt{g}}\sum_{k,s}\frac{\partial}{\partial
  w_k}\left\{\left(\frac{\partial\sqrt{g}H}{\partial q_{lk}} -
    \frac{\partial\sqrt{g} H}{\partial q_{kl}}\right)p^sq_s\right\} 
\end{align}
were shown to be contravariant vectors and constructed in such a way
that terms which contained derivatives of $p^s$ were only contained in
the first two terms on the right hand side of eq.~(\ref{eq:Penergy})
which vanish on stipulation of the field equations. Invoking his
theorem II, Hilbert now showed that $P(\sqrt{g}H)$ could alternatively
be written as
\begin{equation}
  \label{eq:Penergy2}
  P(\sqrt{g}H) = \sum_s \frac{\partial \sqrt{g}Hp^s}{\partial w_s}.
\end{equation}
Hence the expression
\begin{equation}
  \label{eq:energyvector}
  e^l = Hp^l-a^l-b^l-c^l-d^l
\end{equation}
depended linearly on $p^s$ and transformed as a vector under general
coordinate transformations. Invoking the field equations, its
covariant divergence
\begin{equation}
  \label{eq:Edivergence}
  \sum_l \frac{\partial\sqrt{g}e^l}{\partial w_l}=0
\end{equation}
vanishes identically in any coordinate system.%
\footnote{\cite[p.~402]{Hil1915Grundlagen1}.}

Moreover, his other considerations regarding the energy-momentum
tensor remained valid since the electromagnetic part of the energy
followed from his new energy expression as
\begin{equation}
  \label{eq:L_p}
 Lp^l-\sum_k\frac{\partial L}{\partial q_{kl}}p_k-
\frac{1}{2\sqrt{g}}\sum_{k,s}\frac{\partial}{\partial
  w_k}\left\{\left(\frac{\partial\sqrt{g}L}{\partial q_{lk}} -
    \frac{\partial\sqrt{g} L}{\partial q_{kl}}\right)p^sq_s\right\}, 
\end{equation}
as is readily seen by looking at the definition of $e^l$ in
eq.~(\ref{eq:energyvector}) and of $a^l$, $b^l$, $c^l$, $d^l$ in
eqs.~(\ref{eq:a_l})--(\ref{eq:d_l}) and taking into account that the
electromagnetic part $L$ does not depend on the derivatives of the
metric. Using that $L$ depended, in fact, only on the fields $M_{kl}$ and,
invoking the generalized Maxwell equations, Hilbert transformed this
expression to
\begin{equation}
  \label{eq:(18)_pub}
  \sum_{s,k}\left(L\delta^l_s-\frac{\partial L}{\partial M_{lk}}M_{sk} -
\frac{\partial L}{\partial q_l}q_{s}\right)p^s.
\end{equation}
And since eq.~(\ref{eq:(24)}) which followed directly from Hilbert's
second theorem remained valid, it followed that the
electromagnetic part of his new energy concept again yielded Mie's
energy-momentum tensor. Hence the conclusions about Mie's theory as a
special case of Hilbert's theory also remained valid.

But the new energy vector was identically conserved, by virtue of
eq.~(\ref{eq:Edivergence}), for any system of coordinates. Hence the
stipulation of its validity by an extra axiom did not add anything to
the theory, its validity was already implied by the first two axioms,
and hence the third axiom of the proofs was no longer independent from
the first two axioms and had to be dropped. This result was in
agreement with Einstein's conclusions of his final November memoir.

Hilbert might also have thought about the relation of his version of
the field equations given as variational derivatives of an invariant
and Einstein's final field equations which were given in terms of the
Ricci tensor and the Riemann curvature scalar. In the published
version, he added a remark to the effect that for the splitting of $H$
into $H=K+L$, with $K$ representing the Riemann scalar and $L$ the
electromagnetic part depending only on $q_s$, $q_{sl}$ and
$g^{\mu\nu}$, the first term on the left hand side of the field equations
(\ref{eq:(26)}) assumed the explicit form
\begin{equation}
\label{eq:einsteintensor}
  \left[\sqrt{g}K\right]_{\mu\nu} =
  \sqrt{g}(K_{\mu\nu}-\frac{1}{2}Kg_{\mu\nu}),
\end{equation}
and justified this assertion that it followed ``without calculation''
from the fact that $K_{\mu\nu}$ and $g^{\mu\nu}$ are the only tensors
of second rank and $K$ the only invariant which can be formed out of
the $g^{\mu\nu}$ and its first and second derivatives.%
\footnote{\cite[p.~405]{Hil1915Grundlagen1}.}  The argument may not
follow so easily without calculation but is nevertheless true if it is
understood that the second derivatives of the metric tensor enter only
linearly and if the condition is taken into account that the
combination of $K_{\mu\nu}$ and $g^{\mu\nu}K$ has to satisfy the
contracted Bianchi identity (\ref{eq:theorem3}) derived in
Hilbert's Theorem III.%
\footnote{It is also possible that Hilbert took the factor $-1/2$ in
  eq.~(\ref{eq:einsteintensor}) for the trace term $g^{\mu\nu}K$ by
  looking at the field equations of Einstein's final November
  equation, as Corry, Renn and Stachel suggest
  (\cite[p.~1272]{CorryLEtAl1997Decision}). However, Einstein added
  the trace term to the matter tensor $T_{\mu\nu}$ and the equivalence
  of these two forms rests not only on the trivial equivalence of
  putting the trace term to the left or right hand side of the field
  equation but also on the perhaps more problematic identification of
  Einstein's $T_{\mu\nu}$ with Hilbert's
  $\frac{1}{\sqrt{g}}\frac{\partial \sqrt{g}L}{\partial g^{\mu\nu}}$.}

With the representation of his field equations in terms of the Ricci
tensor, Hilbert also had to comment on the relation of his own field
equations to those published by Einstein in his November equations.
The comparison was not so easy because, for one, the source term of
the field equations in Hilbert's and in Einstein's theory were not
necessarily identifiable. In Einstein's theory it was an unspecified
``energy tensor of matter,'' exemplified either by an incoherent,
pressureless flow of particles (dust) or by the energy-momentum
tensor of Maxwellian electrodynamics. For Hilbert it was the
electromagnetic part of his energy vector $e^l$. With Hilbert's
splitting of the ``world function'' $H$ into a gravitational part $K$
and an electromagnetic part $L$ and with his result that, as a
consequence of general covariance, $L$ could only depend on the
electromagnetic potential $q_s$ and on its antisymmetrized derivatives
$M_{sl}$, there is hence some freedom to determine $L$ and hence to
determine the energy-momentum tensor of matter in Mie's sense.
Second, Hilbert never set the determinant of $\sqrt{g}$ equal to 1 as
Einstein was still doing ``in the usual manner''%
\footnote{``in der gewohnten Weise''
  \cite[p.845]{EinsteinA1915Feldgleichungen}.} even in his fourth
November communication where such a specialization was
no longer necessary for the equations to hold.%
\footnote{In a postcard to Schwarzschild, dated 26 February, 1916,
  Hilbert explicitly pointed out this difference to Einstein's theory:
  ``Ich m\"ochte Sie nur darauf aufmerksam machen, dass die Forderung
  Determinante $g=|g_{\mu\nu}|=1$ ganz willk\"urlich und durchaus
  \"uberfl\"ussig ist. In meiner Theorie kommt dieselbe garnicht in
  Frage. Vgl.\ meine Gl.~(21) S.~10.'' (SUB Cod.\ Ms.\ Schwarzschild
  331/8.) The reference is to the field equations (\ref{eq:(26)})
  which in the proofs were eq.~(26). When Klein undertook his
  systematic comparison of Einstein's and Hilbert's energy expressions
  in 1918, he also complained about the changing treatment of the
  $\sqrt{g}$ factor in Einstein's papers,
  \cite[p.~142]{Hil1985EtAlBriefwechsel}.} And the presentation of
what amounted in fact to three different versions of field equations
by Einstein within four weeks may have made it difficult to decide
unambiguously what Einstein's new field equations really were. In
fact, what may seem from hindsight as three distinct and alternative
versions of field equations was presented by Einstein as essentially
the same set of equations plus certain hypotheses about the
choice of coordinates and about the constitution of matter, or modifications
of the vacuum field equations in the presence of matter. Nevertheless,
Hilbert conjectured that the final equations advanced by Einstein were
equivalent to his own if $K$ is taken to be the Riemann scalar:
\begin{quote}
  The resulting differential equations of gravitation are, it seems to
  me, in agreement with the broad theory of general
  relativity established by Einstein in his later papers.%
  \footnote{``Die so zu Stande kommenden Differentialgleichungen der
    Gravitation sind, wie mir scheint, mit der von Einstein in seinen
    sp\"ateren Abhandlungen aufgestellten gro{\ss}z\"ugigen Theorie
    der allgemeinen Relativit\"at im Einklang.''
    \cite[p.~405]{Hil1915Grundlagen1}.}
\end{quote}
And to this passage he added a reference to all four November
communications by Einstein, including the final one, submitted to the
Berlin Academy on November 25 and published a week later on December
2, 1915. With these remarks Hilbert's note was prepared for press, by
mid-February Hilbert received offprints of his paper and, by the end of
March, the last 1915 issue of the \emph{Nachrichten} was eventually
published.

\section{``A certain resentment''}

With and after the publication of their respective notes, neither
Einstein nor Hilbert themselves publicly ever entered into a dispute
of priority.  Nevertheless, during the hectic period of November some
tension between Hilbert and Einstein had arisen as is clear already
from the tone of the correspondence of that time and, in particular,
from Einstein's explicit offer of reconciliation in his response of
December 20 to Hilbert's informing him about his election as a
corresponding member of the G\"ottingen Academy. In this letter
Einstein wrote:
\begin{quote}
There was a certain resentment between us, the cause of which I do not
want to analyse. I have fought against the associated feeling of
bitterness, and with complete success. I again think of you with
unmixed friendliness, and I ask you to try to think of me in the same
way.%
\footnote{``Es ist zwischen uns eine gewisse Verstimmung gewesen,
  deren Ursache ich nicht analysieren will. Gegen das damit verbundene
  Gef\"uhl der Bitterkeit habe ich gek\"ampft, und zwar mit
  vollst\"andigem Erfolge. Ich gedenke Ihrer wieder in ungetr\"ubter
  Freundlichkeit, und bitte Sie, dasselbe bei mir zu versuchen.''
  Einstein to Hilbert, 20 December 1915, CPAE8, Doc.~167.} 
\end{quote}
In fact, Einstein had voiced bitter feelings against Hilbert in a
letter to Heinrich Zangger, one of his closest friends, written on
November 26, one day after Einstein had presented his final field
equations, representing the completion of his General Theory of
Relativity to the Berlin Academy. In this letter to Zangger of
November 26, Einstein had accused Hilbert of the ``nostrification'' 
of his results.%
\footnote{This letter is the subject of \cite{MedicusH1984Comment} and
  is, in fact, undated. It has been dated by context to November 26 in
  Volume 8 of the \emph{Collected Papers of Albert Einstein}.  To be
  precise, Einstein does not mention Hilbert by name but it seems
  beyond doubt that Hilbert was meant.}

As John Earman and Clark Glymour in their account of the November 1915
episode emphasized, ``questions about the priority of discoveries are
often among the least interesting and least important issues in the history
of science.''%
\footnote{\cite[p.~291]{EarmanJEtAl1978Einstein}.}  Nevertheless, for
the reconstruction of conceptual innovation in the natural sciences,
they may have a heuristic value in the identification of new insights,
and Hilbert's first communication on the foundation of physics has
often been commented on in the literature in this respect.%
\footnote{\cite{GuthE1970Contribution}, \cite{MehraJ1974Einstein} seem
  to have been the first to raise the issue of priority.}

Both Hilbert and Einstein saw their achievements of November 1915 as
the culmination of year-long efforts of scientific research along
their respective research programs. This may account for a certain
irritability on both sides. But since Einstein's and Hilbert's
research programs were by no means identical, the question arises as
to the identification of those issues where both scientists felt that
they had to secure their claims of priority. Einstein's resentment
against Hilbert cannot have been induced by Hilbert's publication of
the field equations in the explicit form in terms of the Ricci tensor
and the Riemann scalar, nor in the establishment of full, unrestricted
general covariance. His accusation of nostrification was expressed at
the time of Hilbert's first proofs which did not yet contain the
explicit form of the field equations and which still postulated a
restriction of the covariance by the additional four space-time
equations postulated by the third axiom in the proofs. And this third
axiom was only dropped in the published version when Hilbert had
found his new energy vector. Hence, one may also question whether
even Einstein's offer of reconciliation of December 20 was a reaction to
Hilbert's revisions of his paper.

It seems more probable that Einstein regarded Hilbert's axiomatic
interpretation of his theory as a ``nostrification'' and that he
objected to Hilbert's treating his gravitation theory as a mere
mathematical preliminary to his own theory. In fact, in his first
axiom Hilbert had postulated that the ``world function'' $H$ would
depend on the components of the metric tensor $g^{\mu\nu}$, its first
and second derivatives, as well as on the electromagnetic potential
$q_s$ and its derivatives. And both quantities had been introduced in
the preceding paragraph on the same footing as the
``quantities which characterize the processes in the world,''%
\footnote{``Die das Geschehen in $w_s$ charakterisierenden
  Gr\"o{\ss}en ...'' Proofs, p.~1, \cite[p.~395]{Hil1915Grundlagen1}.}
namely 
\begin{quote}
  1) the ten gravitation potentials $g_{\mu\nu}$ ($\mu,\nu=1,2,3,4$)
  with symmetric tensor character with respect to an arbitrary
  transformation of the world parameter $w_s$;\\
  2) the four electrodynamic potentials $q_s$ with vector character in
  the same sense.%
  \footnote{``1) die zehn Gravitationspotentiale $g_{\mu\nu}$
    ($\mu,\nu=1,2,3,4$) mit symmetrischem Tensorcharakter gegen\"uber
    einer beliebigen Transformation der Weltparameter $w_s$; 2) die
    vier elektrodynamischen Potentiale $q_s$ mit Vektorcharakter im
    selben Sinne.'' Proofs, p.~1}
\end{quote}
From hindsight, it was the introduction of the metric tensor which
implied the most radical break with classical space-time concepts by
providing both the chronogeometrical and the inertio-gravitational
structures and which thus represented \emph{the} ``crucial step in the
development of general relativity.''%
\footnote{\cite[p.~293]{StachelJ1995History}; see also
  \cite{StachelJ1994Changes}.} But already at the time, Einstein had a
clear understanding of the revolutionary and innovative conceptual
implications of his general theory of relativity which, at that time,
had also made him an outsider in the field of gravitation theory.%
\footnote{For a contemporary review of gravitational theories, see
  \cite{AbrahamM1915Gravitationstheorien}.} Hilbert had acknowledged
``the tremendous research problems of Einstein and his perspicaciously
devised methods for their solution''%
\footnote{``[...] die gewaltigen Problemstellungen von Einstein sowie
  dessen scharfsinnige zu ihrer L\"osung ersonnenen Methoden,''
  proofs, p.~1, \cite[p.~395]{Hil1915Grundlagen1}.} in the first
paragraph of his note. But, in the published version, he explicitly gave
credit to Einstein for the introduction of the metric tensor, adding
that those ten gravitational potentials were ``first introduced by Einstein.''%
\footnote{``[...] die zehn zuerst von Einstein eingef\"uhrten
  Gravitationspotentiale [...],'' \cite[p.~395]{Hil1915Grundlagen1}.}

While Hilbert thus gave credit to the conceptual justification of
Einstein's metric theory of gravitation, he nevertheless claimed that
he had independently derived the field equations of General Relativity
from a variational principle. So, of course, did Einstein who did not
refer to Hilbert in his final communication. To be sure, since John
Norton's 1984 account of Einstein's route towards general relativity who
argued that ``Einstein's final steps were self-contained''%
\footnote{\cite[p.~314]{NortonJ1984Einstein}.} no serious argument was
ever advanced disputing Einstein's independence in deriving his field
equations of November 25. Nevertheless, since Hilbert's published note
only contained the date of presentation to the Academy, November 20,
predating Einstein's fourth communication, which was presented to
the Berlin Academy on November 25, by five days, some commentators have
haphazardly conjectured that Einstein may have taken the trace term
introduced in his final communication by looking at Hilbert's paper.%
\footnote{See, e.g., \cite[p.~421]{FoelsingA1993Einstein}.}
This conjecture can now be regarded as definitely refuted by the first
proofs of Hilbert's paper.%
\footnote{\cite{CorryLEtAl1997Decision}.}

But the independence of Einstein's discovery was never a point of
dispute between Einstein and Hilbert. Nor was the independence of
Hilbert's derivation of the field equation ever disputed by Einstein.
Hilbert claimed priority for the introduction of the Riemann scalar into the
action principle and the derivation of the field equations from it,%
\footnote{Thus, Hilbert apparently had objected to the presentation in
  a paper by Herglotz on the geometric implications of the
  introduction of the Riemann tensor into gravitation theory published
  in early 1917 (\cite{HerglotzG1916Gravitationstheorie}). In a
  defensive response, Herglotz admitted that he should have pointed
  out that the tensor $K_{\mu\nu}-\frac{1}{2}g_{\mu\nu}K$ first
  appeared naturally as a variation of $\int K\sqrt{g}dw$ in Hilbert's
  paper. (``Ich h\"atte freilich auf das erstmalige nat\"urliche
  Auftreten des Tensors $K_{\mu\nu}-\frac{1}{2}g_{\mu\nu}K$ als
  Variation von $\int K\sqrt{g}dw$ in Ihren `Grundlagen' besonders hinweisen
  sollen.'' Herglotz to Hilbert, undated, SUB Cod.\ Ms.\ Hilbert 147.)
  
  And in a draft of a letter to Weyl, dated 22 April 1918, written
  after he had read the proofs of the first edition of Weyl's
  ``Raum---Zeit---Materie'' Hilbert also objected to being slighted in
  Weyl's exposition. In this letter again ``in particular the use of
  the Riemannian curvature [scalar] in the Hamiltonian integral''
  (``insbesondere die Verwendung der Riemannschen Kr\"ummung unter dem
  Hamiltonschen Integral'') was claimed as one of his original
  contributions. SUB Cod.\ Ms.\ Hilbert 457/17.} and Einstein admitted
publicly that Hilbert (and Lorentz) had succeeded in giving the
equations of general relativity a particularly lucid form by
deriving them from a single variational principle.%
\footnote{``In letzter Zeit ist es H.A.~Lorentz und D.~Hilbert
  gelungen, der allgemeinen Relativit\"atstheorie dadurch eine
  besonders \"ubersichtliche Gestalt zu geben, da{\ss} sie deren
  Gleichungen aus einem einzigen Variationsprinzipe ableiten.''
  \cite[p.~1111]{EinsteinA1916Prinzip}.}

Hilbert's irritation in November 1915, I would like to suggest,
referred to Einstein's ``Nachtrag'' to his first November
communication, presented to the Berlin Academy on November 11.%
\footnote{\cite{EinsteinA1915Nachtrag}.}  In this ``Nachtrag''
Einstein had advanced generally covariant field equations, equating
the Ricci tensor to the energy-momentum tensor of matter, and
justified these equations by the hypothesis that the trace of the
energy-momentum tensor of matter vanish. This was a necessary
consequence of these field equations if one wanted to fix the
coordinates such that $\sqrt{-g}$ be a constant, and this coordinate
condition had to be imposed in order to recover the field equations of
Einstein's first November communication. Since the trace of the
energy-momentum tensor vanishes for the electromagnetic
energy-momentum tensor but not necessarily for other tensors, e.g. not
for the energy-momentum tensor of incoherent pressureless dust, the
field equations hence implied a hypothesis about the constitution of
matter. As Einstein said:
\begin{quote}
  There are indeed not a few who hope to be able to reduce matter to
  purely electromagnetic processes. These processes, however, would
  have to be governed by a theory which generalizes Maxwell's
  electrodynamics to a completed theory.%
\footnote{``Es gibt sogar nicht wenige, die hoffen, die Materie auf
  rein elektromagnetische Vorg\"ange reduzieren zu k\"onnen, die
  allerdings einer gegen\"uber Maxwell's Elektrodynamik
  vervollst\"andigten Theorie gem\"a{\ss} vor sich gehen w\"urden.''
  \cite[p.~799]{EinsteinA1915Nachtrag}.}
\end{quote}
Clearly, Hilbert must have understood this as a reference to Mie's
electrodynamic field theory of matter. And Einstein's claim that the
field equations implied a hypothesis about the constitution of matter
just touched on the ``Leitmotiv'' of Hilbert's own theory. Indeed, as
was discussed above, Hilbert interpreted his first theorem about the
implied existence of four identities between the 14 field equations
and their derivatives, in the sense that the electrodynamic phenomena
are effects of gravitation. And it is well possible that Hilbert had
informed Einstein about this characteristic of his theory in that
``friendly letter'' which, unfortunately lost, he had sent Einstein in
response to Einstein's initial correspondence of November 7.

But the subsequent introduction of the trace term in Einstein's field
equations of November 25, implied not only that energy-momentum
conservation no longer imposed a restriction on the admissible
coordinates. It also made the hypothesis of the ``Nachtrag''
superfluous, a consequence which Einstein explicitly pointed out:
\begin{quote}
  On the other hand the postulate of general relativity cannot
  disclose anything about the other phenomena of nature that would not
  already follow from special relativity. My former opinion, expressed
  recently in this forum, was erroneous.%
  \footnote{``Dagegen vermag das allgemeine Relativit\"atspostulat uns
    nichts \"uber das Wesen der \"ubrigen Naturvorg\"ange zu
    offenbaren, was nicht schon die spezielle Relativit\"atstheorie
    gelehrt h\"atte.  Meine in dieser Hinsicht neulich an dieser
    Stelle ge\"au{\ss}erte Meinung war irrt\"umlich.''
    \cite[p.~847]{EinsteinA1915Feldgleichungen}.}
\end{quote}
With Einstein's retreat regarding his claim about implications of his
gravitational field equations for an inherent electromagnetic
constitution of matter, the danger of a priority problem for Hilbert,
as far as Einstein was concerned, had vanished. 

\section{Concluding remarks}

The substantial lasting innovation of Hilbert's first note on the
foundation of physics was the foundation of Einstein's general theory
of relativity on an invariant variational principle as an equivalent
representation of the gravitational and electromagnetic field
equations and the mathematical elaboration of some consequences which
follow alone from the invariance of the action integral with respect
to arbitrary transformations of coordinates. These insights include
the discovery of a special case of Noether's second theorem, the
derivation of the generalized contracted Bianchi identities from the
variational principle in his Theorem II, as well as the introduction of the
Riemann curvature scalar into the variational integral. Other
innovative features of his note have not stood the test of time. Among
these are his ideas on a unified field theory of gravitation and
electromagnetism and his energy vector.

With regard to the final establishment of a theory of unrestricted,
general covariance and the interaction between Hilbert and Einstein in
this matter, I should like to venture the following historical
assessment of Hilbert's work. Hilbert's knowledge and understanding of
the calculus of variations and of invariant theory readily put him
into a position to fully grasp Einstein's gravitation theory of 1914
and the mathematical intricacies of the derivation of its field
equations from a variational principle. This understanding and the
subsequent axiomatic reinterpretation of Einstein's theory as well as
Hilbert's way of restricting general covariance in order to guarantuee
energy conservation and causality by means of a third, independent
axiom in the first proofs of his note, posed, it seems to me,
objectively a threat to the delicate, metastable balance between
physical conceptions and their mathematical representation%
\footnote{\cite{RennJEtAl1998Heuristics}.}  of Einstein's
\emph{Entwurf} theory, as achieved by his 1914 exposition of the
\emph{Formale Grundlage}. And whatever Hilbert later learnt from
reading Einstein's final November communication while reviewing the
proofs of his note, I should like to suggest that Hilbert's own
justification of general covariance by means of the revised energy
concept of the published paper, was based also on arguments of
internal mathematical coherency which were independent from Einstein's
considerations.

Hilbert's specific contributions to the history of general relativity
as well as their limitations were conditioned by a vast knowledge of
mathematics and a broad knowledge of contemporary theoretical physics,
by the heuristics of Hilbert's axiomatic method of identifying basic
assumptions and their respective implications and of looking for
fundamental and intricate mathematical questions associated with
physics, as well as by Hilbert's belief in the unity of the
mathematical sciences and in the feasibility of turning physics into a
mathematical discipline based on an axiomatic foundation.

\section*{Acknowledgments}

I wish to thank William Ewald, Gerd Gra{\ss}hoff, Ralf Haubrich, and
Ulrich Majer for many clarifying and encouraging discussions.
Preliminary versions of this paper were presented at the Workshop
``Space-Time, Quantum Entanglement and Critical Epistemology: A
Workshop in Honor of John Stachel,'' Berlin, 5/6 June, 1998, and at
the Workshop ``Wissenschaften im Umbruch. Logik, Mathematik, Physik in
G\"ottingen, 1890--1930,'' G\"ottingen, 12/13 June, 1998.  I wish to
thank J\"urgen Renn and John Stachel for giving me the opportunity to
present this paper in Berlin and for critical comments, and John
Stachel for further discussions at the G\"ottingen workshop. A
critical reading of a previous draft by Hubert Goenner, some useful
comments by Hans-Joachim Dahms, Michael Hallett, Michel Janssen, Skuli
Sigurdsson, and Annette Vogt, help on the English translation of
German quotes by William Ewald, and further critical revision of my
English by Elizabeth Eck greatly helped to improve the present article.

Unpublished manuscripts are quoted by kind permission of the
\emph{Nieders\"achsische Staats-- und Universit\"atsbibliothek}
(\emph{Handschriftenabteilung}) and of the 
\emph{Akademie der Wissenschaften} in G\"ottingen.

\newpage

\section*{Abbreviations}

\begin{tabbing}
xxxxxxxx\=\kill
CPAE:\>The Collected Papers of Albert Einstein\\
GAA: \>Archiv der Akademie der Wissenschaften zu G\"ottingen.\\
GM:  \>Gesch\"aftliche Mitteilungen der K\"oniglichen Gesellschaft\\
     \> der Wissenschaften in G\"ottingen\\
JDM: \>Jahresbericht der Deutschen Mathematikervereinigung. 2.~Abteilung\\
SUB: \>Nieders\"achsische Staats- und Universit\"atsbibliothek G\"ottingen, \\
     \>Handschriftenabteilung.\\

\end{tabbing}

\newcommand{\WkA}{$\langle$}
\newcommand{\WkZ}{$\rangle$}
\newcommand{\Seiten}{pp} %\def\Seiten{S.\@\xspace}
\newcommand{\citeauthoryear}[3]{#2 #3}
\newcommand{\zsp}{{\vrule height0ex width0ex depth0pt}}
%\bibliographystyle{biblio_h}
%\bibliography{paper}

\begin{thebibliography}{}
  
\bibitem[\citeauthoryear{Abraham}{Abraham}{1902a}]{AbrahamM1902Dynamik}
  Max Abraham: Dynamik des Elektrons. --- Nachrichten von der
  Gesellschaft der Wissenschaften zu Göttingen.  {\upshape
    Mathematisch-Physikalische Klasse} (1902), 20--41.
  
\bibitem[\citeauthoryear{Abraham}{Abraham}{1902b}]{AbrahamM1902Prinzipien}
  Max Abraham: Prinzipien der Dynamik des Elektrons.  ---
  Physikalische Zeitschrift 4 (1902), 57--63.
  
\bibitem[\citeauthoryear{Abraham}{Abraham}{1903}]{AbrahamM1903Prinzipien}
  Max Abraham: Prinzipien der Dynamik des Elektrons.  --- Annalen der
  Physik 10 (1903), 105--179.
  
\bibitem[\citeauthoryear{Abraham}{Abraham}{1915}]{AbrahamM1915Gravitationstheorien}
  Max Abraham: Neuere Gravitationstheorien. --- Jahrbuch der
  Radioaktivität und Elektronik 11 (1915), 470--520.
  
\bibitem[\citeauthoryear{Behrens and Hecke}{Behrens and
    Hecke}{1912}]{BehrensWEtAl1912Bewegung} Wilhelm Behrens and Erich
  Hecke: Ueber die geradlinige Bewegung des Bornschen starren
  Elektrons. --- Nachrichten von der Gesellschaft der Wissenschaften
  zu Göttingen. {\upshape Mathematisch-Physikalische Klasse} (1912),
  849--860.
  
\bibitem[\citeauthoryear{Blum}{Blum}{1994}]{BlumP1994Bedeutung} Petra
  Blum: Die Bedeutung von Variationsprinzipien in der Physik für David
  Hilbert (Staatsexamensarbeit). --- Mainz: Universität Mainz 1994.

\bibitem[\citeauthoryear{Blumenthal}{Blumenthal}{1935}]{BlumenthalO193%
    5Lebensgeschichte} Otto Blumenthal: Lebensgeschichte [David
  Hilberts]. --- \upshape In \cite{Hil-3}, 388--429.

\bibitem[\citeauthoryear{Boltzmann}{Boltzmann}{1897}]{BoltzmannL1897Vo%
    rlesungen} Ludwig Boltzmann: Vorlesungen ueber die Principe der
  Mechanik. I.~Theil enthaltend die Principe, bei denen nicht
  Ausdrücke nach der Zeit integrirt werden, welche Variationen der
  Coordinaten oder ihrer Ableitungen nach der Zeit enthalten. ---
  Leipzig: Johann Ambrosius Barth 1897.
  
\bibitem[\citeauthoryear{Born}{Born}{1909a}]{BornM1909Dynamik} Max
  Born: Über die Dynamik des Elektrons in der Kinematik des
  Relativitätsprinzips. --- Physikalische Zeitschrift 10 (1909),
  814--817.
  
\bibitem[\citeauthoryear{Born}{Born}{1909c}]{BornM1909Theorie} Max
  Born: Die Theorie des starren Elektrons in der Kinematik des
  Relativitätsprinzips. --- Annalen der Physik 30 (1909), 1--56, 840.
  
\bibitem[\citeauthoryear{Born}{Born}{1909b}]{BornM1909Masse} Max Born:
  Die träge Masse und das Relativitätsprinzip. --- Annalen der Physik
  28 (1909), 571--584.
  
\bibitem[\citeauthoryear{Born}{Born}{1910}]{BornM1910Kinematik} Max
  Born: Zur Kinematik des starren Körpers im System des
  Relativitätsprinzips. --- Nachrichten von der Gesellschaft der
  Wissenschaften zu Göttingen. {\upshape Mathematisch-Physikalische
    Klasse} (1910), 161--179.
  
\bibitem[\citeauthoryear{Born}{Born}{1914}]{BornM1914Impulsenergiesatz}
  Max Born: Der Impuls-Energie-Satz in der Elektrodynamik von Gustav
  Mie. --- Nachrichten von der Gesellschaft der Wissenschaften zu
  Göttingen. {\upshape Mathematisch-Physikalische Klasse} (1914),
  23--36.
  
\bibitem[\citeauthoryear{Born}{Born}{1922}]{BornM1922Hilbert} Max
  Born: Hilbert und die Physik. --- Die Naturwissenschaften 10 (1922),
  88--93.
  
\bibitem[\citeauthoryear{Abraham Brown and Pippard}{Brown, Abraham and
    Pippard}{1995}]{BrownLEtAl1995Physics} Laurie~M. Brown, Pais
  Abraham and Sir~Brian Pippard (ed.): Twentieth Century Physics,
  Vol.~1. --- Bristol, Philadelphia, New York: Institute of Physics
  Publsihing and American Institute of Physics Press 1995.
  
\bibitem[\citeauthoryear{Fickler Carmeli and Witten}{Carmeli, Fickler
    and Witten}{1970}]{CarmeliMEtAl1970Relativity} Moshe Carmeli,
  Stuart~I. Fickler and Louis Witten (ed.): Relativity. Proceedings of
  the Relativity Conference in the Midwest held at Cincinnati, Ohio,
  June~2--6, 1969. --- New York, London: Plenum Press 1970.
  
\bibitem[\citeauthoryear{Cattani and De~Maria}{Cattani and
    De~Maria}{1989}]{CattaniCEtAl1989Controversy} Carlo Cattani and
  Michelangelo De~Maria: The 1915 Epistolary Controversy between
  Einstein and Tullio Levi-Civita. --- \upshape In
  \cite{HowardDEtAl1989Einstein}, 175--200.
  
\bibitem[\citeauthoryear{Corry}{Corry}{1996}]{CorryL1996HilbertPhysics}
  Leo Corry: Hilbert and Physics (1900--1915).  {\upshape Preprint
    43}. --- Max-Planck-Institut für Wissenschaftsgeschichte (1996).
  
\bibitem[\citeauthoryear{Renn Corry and Stachel}{Corry, Renn and
    Stachel}{1997}]{CorryLEtAl1997Decision} Leo Corry, Jürgen Renn and
  John Stachel: Belated Decision in the Hilbert-Einstein Priority
  Dispute. --- Science 278 (1997), 1270--1273.
  
\bibitem[\citeauthoryear{Corry}{Corry}{1997a}]{CorryL1997Minkowski}
  Leo Corry: Hermann Minkowski and the Postulate of Relativity. ---
  Archive for History of Exact Sciences 51 (1997), 273--314.
  
\bibitem[\citeauthoryear{Corry}{Corry}{1997b}]{CorryL1997Hilbert} Leo
  Corry: David Hilbert and the Axiomatization of Physics
  \WkA{}1894--1905\WkZ{}. --- Archive for History of Exact Sciences 51
  (1997), 83--198.

\bibitem[\citeauthoryear{Corry}{Corry}{1998}]{CorryL1998Hilbert} Leo
  Corry: David Hilbert between Mechanical and Electromagnetic Reductionism
  \WkA{}1910--1915\WkZ{}. --- Archive for History of Exact Sciences
  {\bf 53} (1998).
    
\bibitem[\citeauthoryear{CPAE1}{CPAE1}{1987}]{CPAE1} John Stachel
  (ed.): The Collected Papers of Albert Einstein. Vol.~1: The Early
  Years, 1879-1902. --- Princeton: Princeton University Press 1987.
  
\bibitem[\citeauthoryear{CPAE4}{CPAE4}{1995}]{CPAE4} Martin~J. Klein
  et~al.\zsp (ed.): The Collected Papers of Albert Einstein. Vol.~4:
  The Swiss Years: Writings, 1912--1914. --- Princeton: Princeton
  University Press 1995.
  
\bibitem[\citeauthoryear{CPAE5}{CPAE5}{1993}]{CPAE5} Martin~J. Klein,
  A.J. Kox and Robert Schulmann (ed.): The Collected Papers of Albert
  Einstein. Vol.~5. The Swiss Years: Correspondence, 1902--1914. ---
  Princeton: Princeton University Press 1993.
  
\bibitem[\citeauthoryear{CPAE6}{CPAE6}{1996}]{CPAE6} A.J. Kox,
  Martin~J. Klein and Robert Schulmann (ed.): The Collected Papers of
  Albert Einstein. Vol.~6. The Berlin Years: Writings, 1914--1917. ---
  Princeton: Princeton University Press 1996.

\bibitem[\citeauthoryear{CPAE8}{CPAE8}{1998}]{CPAE8} R. Schulmann
  et~al.\zsp (ed.): The Collected Papers of Albert Einstein. Vol.~8:
  The Berlin Years: Correspondence, 1914--1918. --- Princeton:
  Princeton University Press 1998.
  
\bibitem[\citeauthoryear{Dick}{Dick}{1970}]{DickA1970Noether} Auguste
  Dick: Emmy Noether. 1882--1935. --- Beihefte zur Zeitschrift
  ``Elemente der Mathematik'' 13 (1970), 1--45.
  
\bibitem[\citeauthoryear{Earman and Glymour}{Earman and
    Glymour}{1978}]{EarmanJEtAl1978Einstein} John Earman and Clark
  Glymour: Einstein and Hilbert: Two Months in the History of General
  Relativity. --- Archive for History of Exact Sciences 19 (1978),
  291--308.
  
\bibitem[\citeauthoryear{Earman and Janssen}{Earman and
    Janssen}{1993}]{EarmanJEtAl1993Explanation} John Earman and Michel
  Janssen: Einstein's Explanation of the Motion of Mercury's
  Perihleion. --- \upshape In \cite{EarmanJEtAl1993Attraction},
  129--172.
  
\bibitem[\citeauthoryear{Janssen Earman and Norton}{Earman, Janssen
    and Norton}{1993}]{EarmanJEtAl1993Attraction} John Earman, Michel
  Janssen and John~D. Norton (ed.): The Attraction of Gravitation. New
  Studies in the History of General Relativity. --- Boston, Basel,
  Berlin: Birkhäuser 1993.  \newblock ( Einstein Studies; 5).
  
\bibitem[\citeauthoryear{Einstein and Grossmann}{Einstein and
    Grossmann}{1913}]{EinsteinAEtAl1913Entwurf} Albert Einstein and
  Marcel Grossmann: Entwurf einer verallgemeinerten
  Relativitätstheorie und einer Theorie der Gravitation. --- Leipzig
  und Berlin: B.G.~Teubner 1913.

\bibitem[\citeauthoryear{Einstein}{Einstein}{1914}]{EinsteinA1914Grund%
    lage} Albert Einstein: Die formale Grundlage der allgemeinen
  Relativitätstheorie. --- Sitzungsberichte der Preussischen Akademie
  der Wissenschaften zu Berlin (1914), 1030--1085.
  
\bibitem[\citeauthoryear{Einstein and Grossmann}{Einstein and
    Grossmann}{1914}]{EinsteinAEtAl1914Kovarianzeigenschaften} Albert
  Einstein and Marcel Grossmann: Kovarianzeigenschaften der
  Feldgleichungen der auf die verallgemeinerte Relativitätstheorie
  gegründeten Gravitationstheorie. --- Zeitschrift für Mathematik und
  Physik 63 (1914), 215--225.

\bibitem[\citeauthoryear{Einstein}{Einstein}{1915a}]{EinsteinA1915Feld%
    gleichungen} Albert Einstein: Die Feldgleichungen der Gravitation.
  --- Sitzungsberichte der Preussischen Akademie der Wissenschaften zu
  Berlin (1915), 844--847.

\bibitem[\citeauthoryear{Einstein}{Einstein}{1915d}]{EinsteinA1915Erkl%
    aerung} Albert Einstein: Erklärung der Perihelbewegung des Merkur
  aus der allgemeinen Relativitätstheorie. --- Sitzungsberichte der
  Preussischen Akademie der Wissenschaften zu Berlin (1915), 831--839.

\bibitem[\citeauthoryear{Einstein}{Einstein}{1915b}]{EinsteinA1915Rela%
    tivitaetstheorie} Albert Einstein: Zur allgemeinen
  Relativitätstheorie.  --- Sitzungsberichte der Preussischen Akademie
  der Wissenschaften zu Berlin (1915), 778--786.

\bibitem[\citeauthoryear{Einstein}{Einstein}{1915c}]{EinsteinA1915Nach%
    trag} Albert Einstein: Zur allgemeinen Relativitätstheorie
  (Nachtrag). --- Sitzungsberichte der Preussischen Akademie der
  Wissenschaften zu Berlin (1915), 799--801.

\bibitem[\citeauthoryear{Einstein}{Einstein}{1916a}]{EinsteinA1916Grun%
    dlage} Albert Einstein: Die Grundlage der allgemeinen
  Relativitätstheorie. --- Annalen der Physik 49 (1916), 769--822.

\bibitem[\citeauthoryear{Einstein}{Einstein}{1916b}]{EinsteinA1916Prin%
    zip} Albert Einstein: Hamiltonsches Prinzip und allgemeine
  Relativitätstheorie. --- Sitzungsberichte der Preussischen Akademie
  der Wissenschaften zu Berlin (1916), 1111--1116.
  
\bibitem[\citeauthoryear{Eisenstaedt and Kox}{Eisenstaedt and
    Kox}{1992}]{EisenstaedtEtAl1992Studies} Jean Eisenstaedt and A.J.
  Kox (ed.): Studies in the History of General Relativity. --- Boston,
  Basel, Berlin: Birkhäuser 1992.  \newblock ( Einstein Studies; 3).

\bibitem[\citeauthoryear{Enzyklopädie}{Enzyklopädie}{1921}]{Enzyklopae%
    dieV2} Enzyklopädie der mathematischen Wissenschaften mit
  Einschluss ihrer Anwendungen. Band~V.\ Physik. 2.~Teil. --- Leipzig:
  B.~G.~Teubner 1921.

\bibitem[\citeauthoryear{Festschrift}{Festschrift}{1899}]{Festschrift1%
    899} Festschrift zur Feier der Enthüllung des Gauß-Weber-Denkmals
  in Göttingen.  --- Leipzig: Teubner 1899.

\bibitem[\citeauthoryear{Fölsing}{Fölsing}{1993}]{FoelsingA1993Einstei%
    n} Albrecht Fölsing: Albert Einstein. Eine Biographie. ---
  Frankfurt: Suhrkamp 1993.
  
\bibitem[\citeauthoryear{Goenner et~al.}{Goenner
    et~al.}{1998}]{GoennerHEtAl1998Worlds} Hubert Goenner et~al.\zsp
  (ed.): The Expanding Worlds of General Relativity. ---
  Boston, Basel, Berlin: Birkhäuser 1998 (Einstein Studies 7).

\bibitem[\citeauthoryear{Goldberg}{Goldberg}{1977}]{GoldbergS1977Abrah%
    am} Stanley Goldberg: The Abraham Theory of the Electron: The
  Symbiosis of Experiment and Theory. --- Archive for History of Exact
  Sciences 17 (1977), 7--25.
  
\bibitem[\citeauthoryear{Gould and Cohen}{Gould and
    Cohen}{1994}]{GouldCEtAl1994Artifacts} Carol~C. Gould and
  Robert~S. Cohen (ed.): Artifacts, Representations and Social
  Practice. Essays for Marx Wartofsky. --- Dordrecht, Boston, London:
  Kluwer Academic Publishers 1994.
  
\bibitem[\citeauthoryear{Guth}{Guth}{1970}]{GuthE1970Contribution}
  Eugene Guth: Contribution to the History of Einstein's Geometry as a
  Branch of Physics. --- \upshape In
  \cite{CarmeliMEtAl1970Relativity}, 161--209.

\bibitem[\citeauthoryear{Herglotz}{Herglotz}{1911}]{HerglotzG1911Mecha%
    nik} Gustav Herglotz: Über die Mechanik des deformierbaren Körpers
  vom Standpunkte der Relativitätstheorie. --- Annalen der Physik {\bf
    36} (1911), 493--533.

\bibitem[\citeauthoryear{Herglotz}{Herglotz}{1916}]{HerglotzG1916Gravi%
    tationstheorie} Gustav Herglotz: Zur Einsteinschen
  Gravitationstheorie.  --- Berichte über die Verhandlungen der
  königlich sächsischen Gesellschaft der Wissenschaften zu Leipzig.
  {\upshape Mathematisch-physische Klasse} (1916), 199--203.

\bibitem[\citeauthoryear{Hermann}{Hermann}{1968}]{HermannA1968Einstein%
    Sommerfeld} Armin Hermann (ed.): Albert Einstein/Arnold
  Sommerfeld.  Briefwechsel. --- Basel, Stuttgart: Schwabe \& Co.
  1968.
  
\bibitem[\citeauthoryear{Hertz}{Hertz}{1894}]{HertzH1894Prinzipien}
  Heinrich Hertz: Die Prinzipien der Mechanik. In neuem Zusammenhange
  dargestellt. (Gesammelte Werke.  Band~III.) --- Leipzig: Barth
  \WkA{}Meiner\WkZ{} 1894.
  
\bibitem[\citeauthoryear{Hilbert}{Hilbert}{1895}]{Hil1895Linie} David
  Hilbert: Ueber die gerade Linie als kürzeste Verbindung zweier
  Punkte. --- Mathematische Annalen 46 (1895), 91--96.
  
\bibitem[\citeauthoryear{Hilbert}{Hilbert}{1899}]{Hil1899Grundlagen}
  David Hilbert: Grundlagen der Geometrie. --- \upshape In
  \cite{Festschrift1899}, 92~\Seiten.
  
\bibitem[\citeauthoryear{Hilbert}{Hilbert}{1900}]{Hil1900Probleme}
  David Hilbert: Mathematische Probleme. {\upshape Vortrag, gehalten
    auf dem internationalen Mathematiker-Kongreß zu Paris 1900}. ---
  Nachrichten von der Gesellschaft der Wissenschaften zu
  Göttingen. {\upshape Mathematisch-physikalische Klasse} (1900),
  253--297.

\bibitem[\citeauthoryear{Hilbert}{Hilbert}{1901/02}]{Hil1901/02Problem%
    s} David Hilbert: Mathematical Problems. --- Bulletin of the
  American Mathematical Society 8 (1901/02), 437--479.
  
\bibitem[\citeauthoryear{Hilbert}{Hilbert}{1909}]{Hil1909Minkowski}
  David Hilbert: Hermann Minkowski. {\upshape Gedächtnisrede, gehalten
    in der öffentlichen Sitzung der K.~Gesellschaft der Wissenschaften
    zu Göttingen am 1.~Mai 1909}. --- Nachrichten von der
  Gesellschaft der Wissenschaften zu Göttingen. {\upshape
    Geschäftliche Mitteilungen} (1909), 72--101.
  
\bibitem[\citeauthoryear{Hilbert}{Hilbert}{1912}]{Hil1912Grundzuege}
  David Hilbert: Grundzüge einer allgemeinen Theorie der linearen
  Integralgleichungen. --- Leipzig, Berlin: Teubner 1912.  \newblock (
  Fortschritte der mathematischen Wissenschaften in Monographien;
  Heft~3).
  
\bibitem[\citeauthoryear{Hilbert}{Hilbert}{1915}]{Hil1915Grundlagen1}
  David Hilbert: Die Grundlagen der Physik. {\upshape \WkA{}Erste
    Mitteilung.\WkZ{}}. --- Nachrichten von der Gesellschaft
  der Wissenschaften zu Göttingen. {\upshape
    Mathematisch-physikalische Klasse} (1915), 395--407.
  
\bibitem[\citeauthoryear{Hilbert}{Hilbert}{1917}]{Hil1917Grundlagen2}
  David Hilbert: Die Grundlagen der Physik. {\upshape \WkA{}Zweite
    Mitteilung.\WkZ{}}. --- Nachrichten von der Gesellschaft
  der Wissenschaften zu Göttingen. {\upshape
    Mathematisch-physikalische Klasse} (1917), 53--76.
  
\bibitem[\citeauthoryear{Hilbert}{Hilbert}{1924}]{Hil1924Grundlagen}
  David Hilbert: Die Grundlagen der Physik. --- Mathematische Annalen
  92 (1924), 1--32.
  
\bibitem[\citeauthoryear{Hilbert}{Hilbert}{1935}]{Hil-3} David
  Hilbert: Gesammelte Abhandlungen.  Dritter Band. Analysis,
  Grundlagen der Mathematik, Physik, Verschiedenes. Lebensgeschichte.
  --- Berlin: Springer 1935.
  
\bibitem[\citeauthoryear{Hilbert and Klein}{Hilbert and
    Klein}{1985}]{Hil1985EtAlBriefwechsel} David Hilbert and Felix
  Klein: Der Briefwechsel David Hilbert --- Felix Klein
  \WkA{}1886--1918\WkZ{}. Mit Anmerkungen herausgegeben von Günther
  Frei. --- Göttingen: Vandenhoeck \& Ruprecht 1985.
  
\bibitem[\citeauthoryear{Howard and Stachel}{Howard and
    Stachel}{1989}]{HowardDEtAl1989Einstein} Don Howard and John
  Stachel (ed.): Einstein and the History of General Relativity. ---
  Boston, Basel, Berlin: Birkhäuser 1989.  \newblock ( Einstein
  Studies; 1).
  
\bibitem[\citeauthoryear{Howard and Norton}{Howard and
    Norton}{1993}]{HowardDEtAl1993Labyrinth} Don Howard and John~D.
  Norton: Out of the Labyrinth? Einstein, Hertz, and the Göttingen
  Answer to the Hole Argument. --- \upshape In
  \cite{EarmanJEtAl1993Attraction}, 30--62.
  
\bibitem[\citeauthoryear{Jungnickel and McCormach}{Jungnickel and
    McCormach}{1986}]{JungnickelCEtAl1986Mastery2} Christa Jungnickel
  and Russell McCormach: Intellectual Mastery of Nature.  Theoretical
  Physics from Ohm to Einstein. Vol.~2. The Now Mighty Theoretical
  Physics 1870--1925. --- Chicago and London: The University of
  Chicago Press 1986.
  
\bibitem[\citeauthoryear{Kaufmann}{Kaufmann}{1902}]{KaufmannW1902Masse}
  Walter Kaufmann: Die elektromagnetische Masse des Elektrons. ---
  Physikalische Zeitschrift 4 (1902), 54--57.

\bibitem[\citeauthoryear{Kerschensteiner}{Kerschensteiner}{1885}]{Kers%
    chensteinerG1885Vorlesungen} Georg Kerschensteiner (ed.): Dr. Paul
  Gordan's Vorlesungen über die Invariantentheorie. Erster Band:
  Determinanten. --- Leipzig: Teubner 1885.  \newblock (1).

\bibitem[\citeauthoryear{Kerschensteiner}{Kerschensteiner}{1887}]{Kers%
    chensteinerG1887Vorlesungen} Georg Kerschensteiner (ed.): Paul
  Gordan's Vorlesungen über Invariantentheorie. Zweiter Band: Binäre
  Formen. --- Leipzig: Teubner 1887.  \newblock (2).

\bibitem[\citeauthoryear{Kirchhoff}{Kirchhoff}{1877}]{KirchhoffG1877Vo%
    rlesungen1} Gustav Kirchhoff: Vorlesungen über mathematische
  Physik.  Band~1. Mechanik. --- Leipzig: Teubner 1877.
  
\bibitem[\citeauthoryear{Klein}{Klein}{1917}]{KleinF1917Note} Felix
  Klein: Zu Hilberts erster Note über die Grundlagen der Physik. ---
  Nachrichten von der Gesellschaft der Wissenschaften zu Göttingen.
  {\upshape Mathematisch-Physikalische Klasse} (1917), 469--482.

\bibitem[\citeauthoryear{Klein}{Klein}{1918}]{KleinF1918Differentialge%
    setze} Felix Klein: Über die Differentialgesetze für die Erhaltung
  von Impuls und Energie in der Einsteinschen Gravitationstheorie.
  --- Nachrichten von der Gesellschaft der Wissenschaften zu
  Göttingen. {\upshape Mathematisch-Physikalische Klasse} (1918),
  171--189.
  
\bibitem[\citeauthoryear{Lorentz}{Lorentz}{1909}]{LorentzH1909Theory}
  Hendrik~A. Lorentz: The Theory of Electrons and its Applications to
  the Phenomena of Light and Radiant Heat. --- Leipzig: B.G.~Teubner
  1909.
  
\bibitem[\citeauthoryear{Mach}{Mach}{1889}]{MachE1889Mechanik/2} Ernst
  Mach: Die Mechanik in ihrer Entwickelung historisch-kritisch
  dargestellt. (2.~ed.) --- Leipzig: F.A.~Brockhaus 1889.
  
\bibitem[\citeauthoryear{Medicus}{Medicus}{1984}]{MedicusH1984Comment}
  Heinrich~A. Medicus: A comment on the relations between Einstein and
  Hilbert. --- American Journal of Physics 52 (1984), 206--208.
  
\bibitem[\citeauthoryear{Mehra}{Mehra}{1974}]{MehraJ1974Einstein}
  Jagdish Mehra: Einstein, Hilbert, and The Theory of Gravitation. ---
  Dordrecht, Boston: D.~Reidel Publishing Company 1974.
  
\bibitem[\citeauthoryear{Mie}{Mie}{1912a}]{MieG1912Grundlagen1} Gustav
  Mie: Grundlagen einer Theorie der Materie.  Erste Mitteilung. ---
  Annalen der Physik 37 (1912), 511--534.
  
\bibitem[\citeauthoryear{Mie}{Mie}{1912b}]{MieG1912Grundlagen2} Gustav
  Mie: Grundlagen einer Theorie der Materie.  Zweite Mitteilung. ---
  Annalen der Physik 39 (1912), 1--40.
  
\bibitem[\citeauthoryear{Mie}{Mie}{1913}]{MieG1913Grundlagen3} Gustav
  Mie: Grundlagen einer Theorie der Materie. Dritte Mitteilung.
  Schlu{\ss}. --- Annalen der Physik 40 (1913), 1--66.

\bibitem[\citeauthoryear{Minkowski}{Minkowski}{1973}]{MinkowskiH1973Br%
    iefe} Hermann Minkowski: Briefe an David Hilbert. Mit Beiträgen
  und herausgegeben von L.~Rüdenberg und H.~Zassenhaus.  --- Berlin,
  Heidelberg, New York: Springer 1973.

\bibitem[\citeauthoryear{Noether}{Noether}{1918}]{NoetherE1918Variatio%
    nsprobleme} Emmy Noether: Invariante Variationsprobleme. ---
  Nachrichten von der Gesellschaft der Wissenschaften zu Göttingen.
  {\upshape Mathematisch-Physikalische Klasse} (1918), 235--257.
  
\bibitem[\citeauthoryear{Norton}{Norton}{1984}]{NortonJ1984Einstein}
  John Norton: How Einstein Found His Field Equations.  --- Historical
  Studies in the Physical Sciences 14 (1984), 253--316.
  
\bibitem[\citeauthoryear{Pais}{Pais}{1982}]{PaisA1982Subtle} Abraham
  Pais: `Subtle is the Lord ...' The Science and the Life of Albert
  Einstein. --- Oxford and New York: Clarendon Press and Oxford
  University Press 1982.

\bibitem[\citeauthoryear{Pauli}{Pauli}{1921}]{PauliW1921Relativitaetst%
    heorie} Wolfgang Pauli: Relativitätstheorie. --- \upshape In
  \cite{EnzyklopaedieV2}, 539--775.
  
\bibitem[\citeauthoryear{Pauli}{Pauli}{1979}]{PauliW1979Briefwechsel}
  Armin Hermann, Karl von Meyenn and Victor~F. Weisskopf (ed.):
  Wolfgang Pauli. Wissenschaftlicher Briefwechsel mit Bohr, Einstein,
  Heisenberg u.a. Band I: 1919--1929. --- New York, Heidelberg,
  Berlin: Springer 1979.
  
\bibitem[\citeauthoryear{Reich}{Reich}{1994}]{ReichK1994Entwicklung}
  Karin Reich: Die Entwicklung des Tensorkalküls: vom absoluten
  Differentialkalk\"ul zur Relativit\"atstheorie. --- Basel, Boston,
  Berlin: Birkhäuser 1994.
  
\bibitem[\citeauthoryear{Reid}{Reid}{1970}]{ReidC1970Hilbert}
  Constance Reid: Hilbert. --- Berlin, New York: Springer 1970.
  
\bibitem[\citeauthoryear{Renn and Sauer}{Renn and
    Sauer}{1998}]{RennJEtAl1998Heuristics} Jürgen Renn and Tilman
  Sauer: Heuristics and Mathematical Representation in Einstein's
  Search for a Gravitational Field Equation. --- \upshape In
  \cite{GoennerHEtAl1998Worlds}.
  
\bibitem[\citeauthoryear{Rowe}{Rowe}{1989}]{RoweDE1989Klein} David~E.
  Rowe: Klein, Hilbert, and the Göttingen Mathematical Tradition. ---
  Osiris. {\upshape Second Series} 5 (1989), 186--213.

\bibitem[\citeauthoryear{Runge}{Runge}{1949}]{RungeI1949Runge} Iris
  Runge: Carl Runge und sein wissenschaftliches Werk. --- G\"ottingen:
  Vandenhoeck \& Ruprecht 1949.
  
\bibitem[\citeauthoryear{Stachel}{Stachel}{1980}]{StachelJ1980Search}
  John Stachel: Einstein's Search for General Covariance, 1912--1915.
  Revised version of a Paper presented at the Ninth International
  Conference on General Relativity and Gravitation, Jena, 14--19 July
  1980. --- \upshape In \cite{HowardDEtAl1989Einstein}, 63--100.

\bibitem[\citeauthoryear{Stachel}{Stachel}{1992}]{StachelJ1992CauchyPr%
    oblem} John Stachel: The Cauchy Problem in General
  Relativity---The Early Years. --- \upshape In
  \cite{EisenstaedtEtAl1992Studies}, 407--418.
  
\bibitem[\citeauthoryear{Stachel}{Stachel}{1994}]{StachelJ1994Changes}
  John Stachel: Changes in the Concepts of Space and Time brought
  about by Relativity. --- \upshape In \cite{GouldCEtAl1994Artifacts},
  141--162.
  
\bibitem[\citeauthoryear{Stachel}{Stachel}{1995}]{StachelJ1995History}
  John Stachel: History of Relativity. --- \upshape In
  \cite{BrownLEtAl1995Physics}, 249--356.

\bibitem[\citeauthoryear{Thiele}{Thiele}{1997}]{ThieleR1997Variationsr%
    echnung} Rüdiger Thiele: Über die Variationsrechnung in Hilberts
  Werken zur Analysis. --- Zeitschrift für Geschichte der
  Naturwissenschaften, Technik und Medizin (NTM) 5 (1997), 23--42.

\bibitem[\citeauthoryear{Tollmien}{Tollmien}{1991}]{TollmienC1991Habil%
    itation} Cordula Tollmien: Die Habilitation von Emmy Noether an
  der Universität Göttingen. --- Zeitschrift für Geschichte der
  Naturwissenschaften, Technik und Medizin (NTM) 28 (1991), 13--32.

\bibitem[\citeauthoryear{Verzeichnis}{Verzeichnis}{1915}]{Verzeichnis1%
    915WSVorlesungen} Verzeichnis der Vorlesungen auf der
  Georg-Universität zu Göttingen während des Winterhalbjahres 1915/16.
  --- Göttingen: Dieterich'sche Universitätsdruckerei 1915.
  
\bibitem[\citeauthoryear{Vizgin}{Vizgin}{1994}]{VizginV1994Theories}
  Vladimir~P. Vizgin: Unified Field Theories in the First Third of the
  20th Century. --- Basel, Boston, Berlin: Birkhäuser 1994.

\bibitem[\citeauthoryear{Volkmann}{Volkmann}{1900}]{VolkmannP1900Einfu%
    ehrung} Paul Volkmann: Einführung in das Studium der theoretischen
  Physik insbesondere in das der analytischen Mechanik mit einer
  Einführung in die Theorie der physikalischen Erkenntniss. ---
  Leipzig: B.G.~Teubner 1900.
  
\bibitem[\citeauthoryear{Walter}{Walter}{1998}]{WalterS1998Minkowski}
  Scott Walter: Minkowski, Mathematicians and the Mathematical Theory
  of Relativity. --- \upshape In \cite{GoennerHEtAl1998Worlds}.
  
\bibitem[\citeauthoryear{Weyl}{Weyl}{1944}]{WeylH1944Hilbert} Hermann
  Weyl: David Hilbert and his mathematical work. --- Bulletin of the
  American Mathematical Society 50 (1944), 612--654.

\bibitem[\citeauthoryear{Wiechert}{Wiechert}{1899}]{WiechertE1899Grund%
    lagen} Emil Wiechert: Grundlagen der Elektrodynamik. --- \upshape
  In \cite{Festschrift1899}, 112~\Seiten.

\end{thebibliography}
%     \input{bibliogr}

\end{document}